\documentclass[
journal=acsphoton,
manuscript=article,
layout=twocolumn
]{achemso}

\usepackage{graphicx}
\usepackage{mciteplus}
\usepackage{siunitx}
\DeclareSIUnit\angstrom{\protect \text {Å}}
\usepackage{floatflt,epsfig} 
\usepackage{float}
\usepackage{mathtools}
\usepackage{color}
\usepackage{braket}
\usepackage{hyperref}
\usepackage[normalem]{ulem}
\usepackage{pdfpages}
\usepackage{natbib}
\usepackage[english]{babel}
\usepackage{blindtext}
\usepackage{lipsum}  
\usepackage{comment}
\usepackage{amsthm}
\usepackage{tabularx}
\usepackage{bm}
\usepackage{dsfont}
\usepackage{bbold}
\usepackage{ulem}
\usepackage{mwe,tikz}
\usepackage[percent]{overpic}
\usepackage{todonotes}
\usepackage{algorithm}
\usepackage{algpseudocode}
\usepackage{uri}

\DeclareUnicodeCharacter{0301}{\"{o}}

\DeclareSIUnit{\atomicunit}{a.u.}
\DeclareSIUnit{\nanometerscube}{nm$^3$}

    \makeatletter
    \def\@fnsymbol#1{\ensuremath{\ifcase#1\or *\or*\or  \ddagger\or
   \mathsection\or \mathparagraph\or \|\or **\or \dagger\dagger
   \or \ddagger\ddagger \else\@ctrerr\fi}}
    \makeatother

\title{Strong coupling electron-photon dynamics: a real-time investigation of energy redistribution in molecular polaritons}

\author{Matteo Castagnola}
\affiliation{Department of Chemistry, Norwegian University of Science and Technology, 7491 Trondheim, Norway}

\author{Marcus T. Lexander}
\affiliation{Department of Chemistry, Norwegian University of Science and Technology, 7491 Trondheim, Norway}

\author{Enrico Ronca}
\email{enrico.ronca@unipg.it}
\affiliation{Dipartimento di Chimica, Biologia e Biotecnologie, Università degli Studi di Perugia, Via Elce di Sotto, 8,06123, Perugia, Italy}

\author{Henrik Koch}
\email{henrik.koch@ntnu.no}
\affiliation{Department of Chemistry, Norwegian University of Science and Technology, 7491 Trondheim, Norway}

\begin{document}

\begin{abstract}
We analyze the real-time electron-photon dynamics in long-range polariton-mediated energy transfer using a real-time quantum electrodynamics coupled cluster (RT-QED-CC) model, which allows for spatial and temporal visualization of transport processes.
We compute the time evolution of photonic and molecular observables, such as the dipole moment and the photon coordinate, following the excitation of the system induced by short laser pulses.
Our simulation highlights the different time scales of electrons and photons under light-matter strong coupling, the role of dark states, and the differences with the electronic (F{\"o}rster and Dexter) energy exchange mechanisms.
The developed method can simulate multiple high-intensity laser pulses while explicitly retaining electronic and electron-photon correlation and is thus suited for nonlinear optics and transient absorption spectroscopies of molecular polaritons.
\end{abstract}

\maketitle


\section{Introduction}

When a molecular excitation strongly interacts with a confined optical mode (e.g., in a Fabry-Pérot cavity), hybrid light-matter states, called polaritons, are formed.
Polaritons have recently attracted interest from chemists due to pioneering experiments proving modifications in chemical properties, such as photochemistry and ground state reactivity.\cite{hutchison2012modifying, munkhbat2018suppression, thomas2016ground, lather2019cavity, thomas2019tilting, canaguier2013thermodynamics, sau2021modifying, hirai2020modulation, ahn2023modification} 
Several works suggest a modified electronic dynamics under light-matter strong coupling, showing e.g. changes in lifetimes,\cite{wang2018dynamics, hutchison2012modifying} selection rules in multi-photon absorption,\cite{wang2023study} energy transfer,\cite{georgiou2021ultralong, zhong2017energy, zhong2016non, cargioli2024active, coles2014polariton, schafer2019modification, xiang2020intermolecular}  as well as enhanced nonlinear optical properties.\cite{wang2021large, ge2021strongly, chervy2016high, malave2022real, xiang2019manipulating}
The proper understanding of such experimental findings in cavity-modified chemistry thus requires a microscopic model of the electron-photon interplay.

In this paper, we take another step forward in describing polaritons by explicitly modeling laser-driven molecules in quantum cavities via real-time quantum electrodynamics coupled cluster (RT-QED-CC).
The \textit{ab initio} QED-CC parametrization of the polaritonic wave function, which includes electron-electron and electron-photon correlation,\cite{haugland2020coupled, helgaker2013molecular} provides a nonperturbative description of the electron-photon dynamics.
We can hence reliably describe both the long-range photonic correlation and the short-range intra- and intermolecular electronic forces. 
The temporal evolution of the system can be monitored through matter and photon quantities such as the electronic density and the photon number.
The RT-QED-CC method is therefore suited for studying the interaction of molecules and aggregates with confined optical modes from the weak to the ultrastrong coupling regime, even under external high-intensity and ultrashort electric pulses (linear and nonlinear excitation regimes).
Our method can support the study of quantum light spectroscopies,\cite{dorfman2016nonlinear, schlawin2013two, ruggenthaler2018quantum, raymer2021entangled, schlawin2018entangled, rezus2012single, raimond2001manipulating}  high-order harmonic generation,\cite{wang2021large, ge2021strongly, chervy2016high} multi-photon and transient (pump-probe) spectroscopy of molecular polaritons.\cite{gu2023cavity, delpo2020polariton, wang2023study, wang2018dynamics}
We here focus on photon-mediated energy transfer as a clear example of experimentally investigated electron-photon dynamics\cite{georgiou2021ultralong, zhong2017energy, zhong2016non, cargioli2024active, coles2014polariton} with potential applications in light-harvesting systems.

Our method allows spatial and temporal visualization of the transport process, following the real-time quantum coherent oscillations in the molecular density and observables such as the dipole moment.
Moreover, since we explicitly include the QED field, we describe in real-time the role of the optic environment and its entanglement with matter, computing photon quantities such as the photon coordinate.
Our results highlight different time scales in the electron-photon dynamics, the role of the dark states, and the modified interaction of polaritons with external pulses.
Moreover, RT-QED-CC can also accurately model the short-range electronic energy transfer (F{\"o}rster and Dexter), offering a clear discussion of the similarities and differences with the photonic channel.
While our method is so far designed for electronic strong coupling (ESC), we believe the concepts we developed for polaritonic energy redistribution can also be applied to vibrational strong coupling (VSC).\cite{xiang2020intermolecular, cao2022generalized, tibben2023molecular, li2021collective}\\

The paper is organized as follows. 
First, we briefly introduce the theory behind \textit{ab initio} quantum electrodynamics (QED) and describe the QED Hartree-Fock (QED-HF) and QED-CC methods.
Next, we provide a detailed description of the real-time QED-CC wave function and discuss the novel information we can extract by explicitly modeling the photonic degrees of freedom.
We then apply the RT-QED-CC method to study energy transfers in molecular polaritons.
Finally, we summarize the main results presented in the paper and discuss future perspectives and applications of the developed framework.

\section{\textit{Ab initio} coupled cluster for molecular polaritons}\label{sec: theory}

In this section, we introduce the QED-HF and QED-CC methods and focus on the real-time description of the QED-CC parametrization.
The photons are described at the same level as the electrons in a polaritonic wave function.
To this end, we employ the non-relativistic Pauli-Fierz Hamiltonian in the dipole approximation, Born-Oppenheimer approximation, and length representation,\cite{ruggenthaler2023understanding, castagnola2024polaritonic} here expressed in second quantization and atomic units
\begin{align}\label{eq: dipole Hamiltonian}
    H&=\sum_{pq}h_{pq}E_{pq}+\frac{1}{2}\sum_{pqrs}g_{pqrs}e_{pqrs}+h_{nuc}\nonumber\\
    &+\sqrt{\frac{\omega}{2}}(\bm{\lambda}\cdot \bm{d})_{pq}E_{pq}(b^\dagger+b) \nonumber \\
    &+\frac{1}{2}\sum_{pqrs}(\bm{\lambda}\cdot \bm{d})_{pq}(\bm{\lambda}\cdot \bm{d})_{rs}E_{pq}E_{rs}\nonumber\\
    &+\omega {b}^\dagger{b},
\end{align}
where $E_{pq}=\sum_\sigma a^\dagger_{p\sigma}a_{q\sigma}$ and $e_{pqrs}=E_{pq}E_{rs}-\delta_{qr}E_{ps}$ are the spin-adapted singlet electronic operators in second quantization, where the indices $p,q,r$, and $s$ label the one-electron basis.\cite{helgaker2013molecular}
The quantities $h_{pq}$, $g_{pqrs}$, and $\bm{d}_{pq}$ are the one-electron, two-electron, and dipole integrals, respectively. The operators $b^\dagger$ and $b$ create and annihilate photons of a single effective photon mode of frequency $\omega$.
Finally, $\bm{\lambda}$ is the light-matter coupling strength of the photon field.
The Hamiltonian in \autoref{eq: dipole Hamiltonian} is employed for the \textit{ab initio} QED-HF and QED-CC methods,\cite{haugland2020coupled} which we briefly describe in the following sections.

\subsection{QED-HF}
The QED-HF wave function is the direct product of a Slater determinant and a photonic state
\begin{equation}
    \ket{\mathrm{R}}=\ket{\mathrm {HF}}\otimes \sum_{{n}}(b^\dagger)^{n}\ket{0}c_{{n}},
\end{equation}
where the optimal parameters are obtained from the variational principle by minimizing the expectation value of the Hamiltonian in \autoref{eq: dipole Hamiltonian}.\cite{haugland2020coupled}
The orbitals are obtained by diagonalization of the QED-HF Fock matrix, while the photon state is a coherent state determined by the molecular dipole moment\cite{haugland2020coupled}
\begin{align}\label{eq: QEDHF state}
    &\ket{\mathrm R}=\ket{\mathrm{HF}}\otimes U_{\text{QED-HF}}\ket{0}\equiv U_{\text{QED-HF}}\ket{\mathrm{HF},0},
\end{align}
where
\begin{align}\label{eq: coherent-transf}
    &U_{\text{QED-HF}} = \text{exp}\bigg(-\frac{\bm{\lambda}\cdot \braket{\bm{d}}_{\text{QED-HF}}}{\sqrt{2\omega}}(b^\dagger - b)\bigg).
\end{align}

\subsection{QED-CC}

The QED-HF wave function of \autoref{eq: QEDHF state} provides the reference state of the QED-CC parametrization, and the coherent-state transformation is handled via a picture change of the Hamiltonian.\cite{haugland2020coupled, castagnola2024polaritonic}
The QED-CC wave function is thus defined in the QED-HF coherent-state representation $\Tilde{H}=U^\dagger_{\text{QED-HF}}H\,U_{\text{QED-HF}}$, that is, the Hamiltonian used to determine the CC amplitudes is transformed using \autoref{eq: coherent-transf}
\begin{align}\label{eq:QEDHF-transf-Ham}
    \Tilde{H}&=\sum_{pq}h_{pq}E_{pq}+\frac{1}{2}\sum_{pqrs}g_{pqrs}e_{pqrs}+h_{nuc}\nonumber\\
    &+\sqrt{\frac{\omega}{2}}(\bm{\lambda}\cdot (\bm{d}-\braket{\bm{d}}_{\text{QED-HF}}))(b^\dagger+b)\nonumber\\
    &+\frac{1}{2}(\bm{\lambda}\cdot (\bm{d}-\braket{\bm{d}}_{\text{QED-HF}}))^2\nonumber\\
    &+\omega{b}^\dagger{b}.
\end{align}
As for the electronic coupled cluster,\cite{helgaker2013molecular} the QED-CC wave function is given by an exponential parametrization 
\begin{align}
    \ket{\text{QED-CC}}&=e^T\ket{\text{HF},0},\label{eq:QEDCC ansatz}
\end{align}
where the cluster operator $T$ includes electronic, photonic, and electron-photon excitations.\cite{haugland2020coupled}
In this paper, we use the QED-CCSD-1 model,\cite{haugland2020coupled} which includes single and double excitations in the electronic space and a single excitation for the photon
\begin{align}
    T&=T_e+T_{int}+T_p,
\end{align}
where
\begin{align}
    &T_e= \sum_{ai}t_{ai}E_{ai}+\frac{1}{2}\sum_{aibj}t_{aibj}E_{ai}E_{bj}\\
    &T_{int} = \sum_{ai}s_{ai}E_{ai}b^\dagger + 
    \frac{1}{2}\sum_{aibj}s_{aibj}E_{ai}E_{bj}b^\dagger\\
    &T_p = \gamma b^\dagger.
\end{align}
In these equations, $i$ and $j$ label occupied orbitals, and $a$ and $b$ label virtual orbitals of the QED-HF reference determinant.
The QED-CC amplitudes are then obtained by projection of the Schr{\"o}dinger equation.
In the same way, the QED-CC dual state is defined as
\begin{equation}
    \bra{\Lambda}=\bra{\text{HF},0}+\sum_{\mu,n}\bar{t}_{\mu {{n}}}\bra{\mu,{{n}}}e^{-T},
\end{equation}
where $\bar{t}_{\mu {n}}$ are the Lagrangian multipliers for the $\ket{\mu,n}$ excitation.\cite{haugland2020coupled, helgaker2013molecular}

The QED-CC parametrization allows for a flexible description of the ground state, including both electron-electron and electron-photon correlation.
Following the same approach as for electronic CC, we can then obtain information on the molecular-polariton excited states using the equation of motion (EOM) or response formalism.\cite{haugland2020coupled, helgaker2013molecular, pedersen1997coupled, koch1990coupled, castagnola2024polaritonic}
These effectively time-independent approaches rely upon the Fourier transformation of any perturbation applied to the system.
In the following, we describe a different approach based on a real-time propagation of the CC amplitudes.

\subsection{Real-time QED-CC}
We aim to study the time evolution of the electron-photon system subject to an external field $V(t)$ such as an electric pump.
The total Hamiltonian is thus $\tilde H + \Tilde V(t)$ (in the QED-HF coherent state representation of \autoref{eq:QEDHF-transf-Ham}), and we parametrize the time-evolution using time-dependent amplitudes $T(t)$ and a complex global-phase parameter $\alpha(t)$, similarly to electronic TD-CC\cite{sverdrup2023time, skeidsvoll2020time, skeidsvoll2022simulating, nascimento2019general, pedersen2019symplectic, huber2011explicitly}
\begin{equation}\label{eq: RT-QED-CC wave function}
    \ket{\text{QED-CC}}(t)=e^{T(t)}\ket{\text{HF},0}e^{i\alpha(t)}.
\end{equation}
The amplitudes are obtained by projection of the time-dependent Schr{\"o}dinger equation\cite{koch1990coupled, pedersen1997coupled, aurbakken2023time}
\begin{align}
    \frac{d\alpha}{dt}&=-\braket{\text{HF},0|(\tilde H+\tilde V(t))e^{T(t)}|\text{HF},0}\label{eq:epsilon_CC_resp}\\
    \frac{dt_\mu}{dt}&=-i\braket{\mu,0|e^{-T(t)}(\tilde H+\tilde V(t))e^{T(t)}|\text{HF},0}\label{eq:electronic_CC_resp}\\
    \frac{d\gamma}{dt}&=-i\braket{\text{HF},{{n}}|e^{-T(t)}(\tilde H+\tilde V(t))e^{T(t)}|\text{HF},0}\label{eq:photon_CC_resp}\\
    \frac{ds_{\mu}}{dt}&=-i\braket{\mu,{{n}}|e^{-T(t)}(\tilde H+\tilde V(t))e^{T(t)}|\text{HF},0}\label{eq:interaction_CC_resp},
\end{align}
and analogous equations hold for the QED-CC dual state (further details are provided in the Supporting Information).

While response theory would rely on a perturbative expansion to obtain linear and nonlinear response functions in the frequency domain,\cite{castagnola2024polaritonic, koch1990coupled, pedersen1997coupled} we numerically propagate the amplitudes and multipliers in time.
We thus study the electron-photon dynamics computing the mean values of observables $\braket{A}$ at each time step
\begin{equation}
    \braket{ A}(t) = \braket{\Lambda (t) | A|\text{QED-CC}(t)}.\label{eq:mean value}
\end{equation}
The information in the frequency domain can then be obtained via a Fourier transform of \autoref{eq:mean value}.
Notice that in QED, we have access to additional observables from the explicit modeling of the photon field.
We can thus describe photon quantities such as the photon coordinate $q = \frac{b+b^\dagger}{\sqrt{2\omega}}$, which is connected to the electric field of the photons in the cavity, or the photon number $N_{ph}=b^\dagger b$.
We must then account for the picture change of the QED-HF coherent state transformation in \autoref{eq: coherent-transf} since the QED-CC wave function is defined for the Hamiltonian in \autoref{eq:QEDHF-transf-Ham}.\cite{castagnola2024polaritonic}

The RT-QED-CC approach allows for the description of the electron-photon dynamics under ultrashort and intense \textit{classical} electric fields
\begin{equation}\label{eq: electric pulse}
    V(t) = \bm{d}\cdot \bm{E}_{ext}(t) = \bm{d}\cdot \mathcal{E}_0(t) \;\cos(\omega_{ext} t + \varphi),
\end{equation}
where $\mathcal{E}_0(t)$ is an envelope function that describes the shape of the external pulse, and the field-molecule interaction is in the length gauge and dipole approximation.
Moreover, the explicit modeling of the photon field provides additional flexibility in the operator $V(t)$.
We can, in principle,
study the dynamics of the electrons interacting with \textit{quantum} light, for instance, a coherent state, by changing the parameters in the transformation \autoref{eq: coherent-transf}.
The interaction of molecules with a photon number state can be modeled by setting the appropriate initial state conditions for the QED-CC states.
Moreover, while standard molecular spectroscopy focuses on the molecular degrees of freedom, in a QED framework, we can also study the effect of perturbations that couple directly to the photon field (for instance, via external currents), such as
\begin{align}
    V(t) &= \mathcal{A}_0 (b+b^\dagger)\\
    V(t) &= i\mathcal{B}_0 (b-b^\dagger),
\end{align}
and subsequently follow the evolution of molecular observables.

The RT-QED-CC method thus allows for a nonperturbative coherent and correlated time-dependent description of the interaction of photons and molecules, permitting the modeling of transient spectroscopies for molecular polaritons, quantum light spectroscopies, molecules in laser fields, and electron-photon quantum entanglement.\cite{dorfman2016nonlinear, raimond2001manipulating, schlawin2017entangled, rezus2012single, ruggenthaler2018quantum, wang2018dynamics, wang2023study}

\section{Results and discussions}\label{sec: results}

In this section, the RT-QED-CCSD-1 method is employed to study photon-mediated energy transfers. 
The RT-QED-CCSD-1 equations have been implemented in a private branch of the $e^\mathcal{T}$ program,\cite{folkestad2020t} and the computational and implementation details are provided in the Supporting Information.

\subsection{Intermolecular energy transfer}

Electronic intermolecular energy transfer has been experimentally investigated\cite{georgiou2021ultralong, zhong2017energy, zhong2016non, cargioli2024active, coles2014polariton} due to its potential applications in technologies for e.g. solar cells and light-harvesting systems. 
While theoretical models were so far limited to simplified methods,\cite{schafer2019modification} we here provide an \textit{ab initio} QED-CC simulation in real-time, with microscopic spatial and temporal visualization and a correlated description of electronic and photonic quantities.

\begin{figure*}
    \centering
    \includegraphics[width=\textwidth]{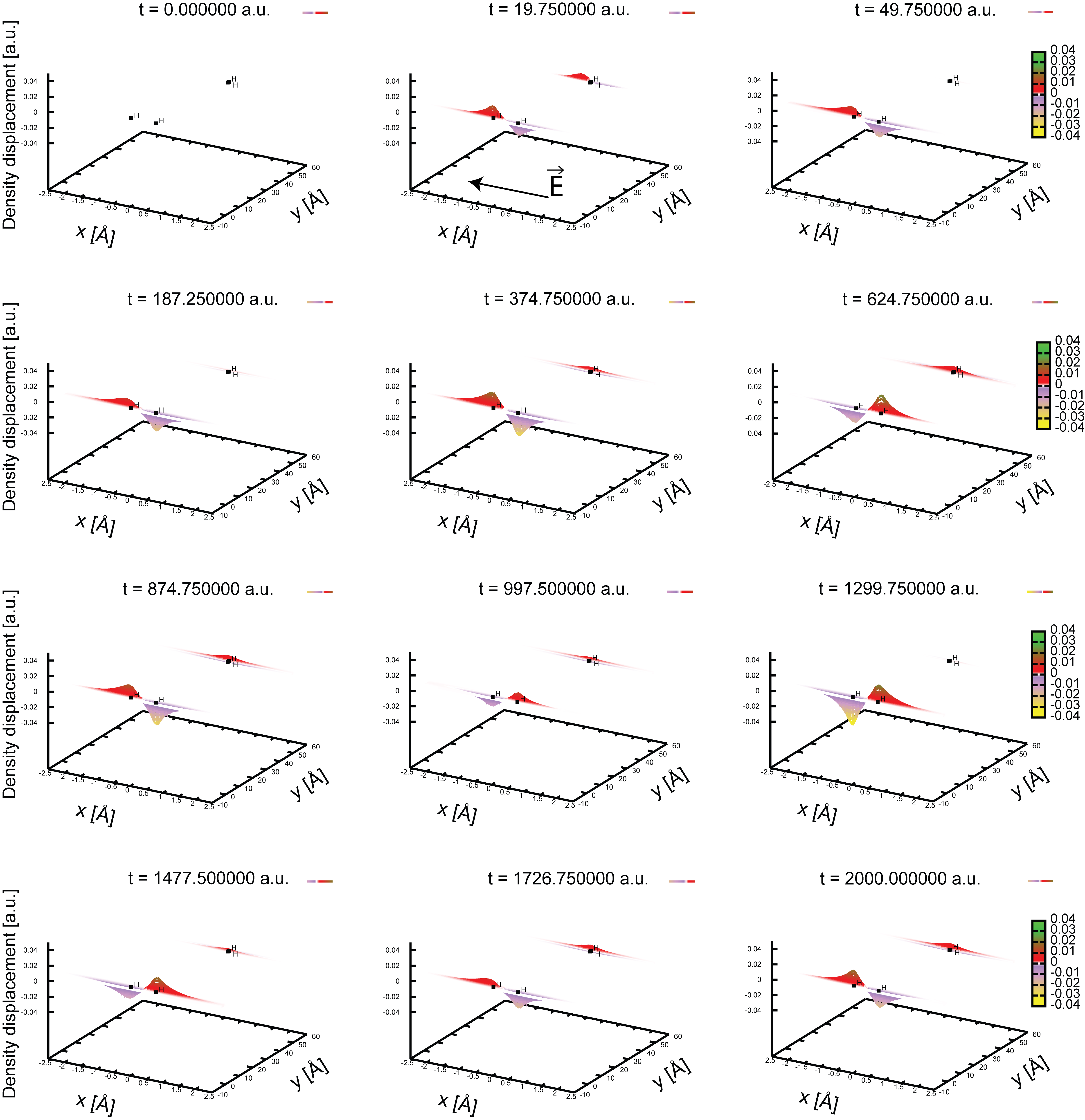}
    \caption{Electron density displacement (compared to the molecular ground state) for two identical but perpendicular $\text H_2$ molecules placed on the xy plane, at different times.
    The distance between them is $D=$ \qty[mode = text]{50}{\angstrom} and the light-matter coupling strength is \qty[mode = text]{0.01}{\atomicunit}, with photon polarization oriented such that both molecules are coupled to the cavity $\bm \epsilon = (1/\sqrt{2}, 1/\sqrt{2}, 0)$.
    The molecules are perturbed by an ultrashort classical electric pulse centered at \qty[mode = text]{20}{\atomicunit} (second panel), but only the $\text H_2$ parallel to the classical electric field is excited after the pulse passes.
    The density of the excited molecule then oscillates in time.
    However, in a quantum cavity, the excitation is transferred to the photon field, which in turn excites the other $\text H_2$, thus allowing for ultralong range energy transfer.
    As time passes, the second molecule transfers its energy back to the photon field and eventually to the first molecule since no decoherence is present so far in the simulations.
    }
    \label{fig:(H2)2 ET}
\end{figure*}

As a simple system to study the long-range polaritonic energy transfer, we consider two identical but perpendicular $\text H_2$ molecules in the xy-plane, with the cavity field tuned to their first bright molecular excitation (see \autoref{fig:(H2)2 ET}).
The light-matter coupling strength is here set to \qty[mode = text]{0.01}{\atomicunit} with polarization $\bm \epsilon = (1/\sqrt{2}, 1/\sqrt{2}, 0)$, such that both the $\text H_2$ are coupled to the optical environment.
The intermolecular distance is large enough to suppress any electronic coupling between them ($D=$ \qty[mode = text]{50}{\angstrom}).
Further computational details are provided in the Supporting Information.
In \autoref{fig:(H2)2 ET}, we show selected snapshots of the differential electronic density  (compared to the ground state) in the xy-plane following the interaction with an ultrashort classical electric pulse (see \autoref{eq: electric pulse}) centered at $t = $ \qty[mode = text]{20}{\atomicunit} and polarized along the x-direction (parallel to one of the $\text H_2$).\footnote{Videos of the energy-transfer dynamics are available in the following repository: \doi{10.5281/zenodo.10813660}.}
While the electric field polarizes both molecules (see the second panel of \autoref{fig:(H2)2 ET}), only one of them is excited after the pulse has passed because of their orientation.
In the absence of the photon coupling, the electron density of the excited molecule oscillates in time while nothing would happen to the second $\text H_2$.
However, when the system is coherently coupled to an optical environment such as a  Fabry-Pérot cavity, the excited molecule can transfer energy to the photon field, which, in turn, excites the other $\text H_2$ molecule.
In \autoref{fig:(H2)2 ET}, we thus see an ultralong energy transfer mediated by the cavity field. 
For computational reasons, we used a specific orientation to selectively excite only one $\text H_2$.
However, longer pulses can be used to selectively excite a single species and thus trigger a cavity-mediated energy transfer between different molecules in quantum cavities.\cite{georgiou2021ultralong, zhong2017energy, zhong2016non, cargioli2024active,coles2014polariton}
Notice that since no decoherence is introduced in the simulation, the excitation will be transferred back and forth between the $\text H_2$ molecules in a quasi-periodic fashion.

The electron-photon dynamics is more easily analyzed by studying the time-evolution of observables such as the dipole moment $\bm d$ and the photon coordinate $q= {(b^\dagger + b}{)}/{\sqrt{2\omega}}$.
In the upper panel of \autoref{fig:(H2)2 50AA, 0.76}, we plot the x and y components of the dipole moment for the process pictorially illustrated in \autoref{fig:(H2)2 ET}.
The y component of the dipole moment provides a simple measure of the excitation of the second $\text H_2$.
The panel clearly illustrates the fast density fluctuations and the slower energy transfer between the two molecules.
The time evolution of the dipole moment for the undressed (out-of-cavity) electron dynamics is reported in the Supporting Information, and it does not show any energy transfer due to the large distance between the molecules. 
To follow the excitation of the quantum cavity, we report the photon coordinate $ \braket{q}={\braket{b^\dagger + b}}/{\sqrt{2\omega}}$ in the lower panel of \autoref{fig:(H2)2 50AA, 0.76}.
\begin{figure}
    \centering
    \includegraphics[width=.45\textwidth]{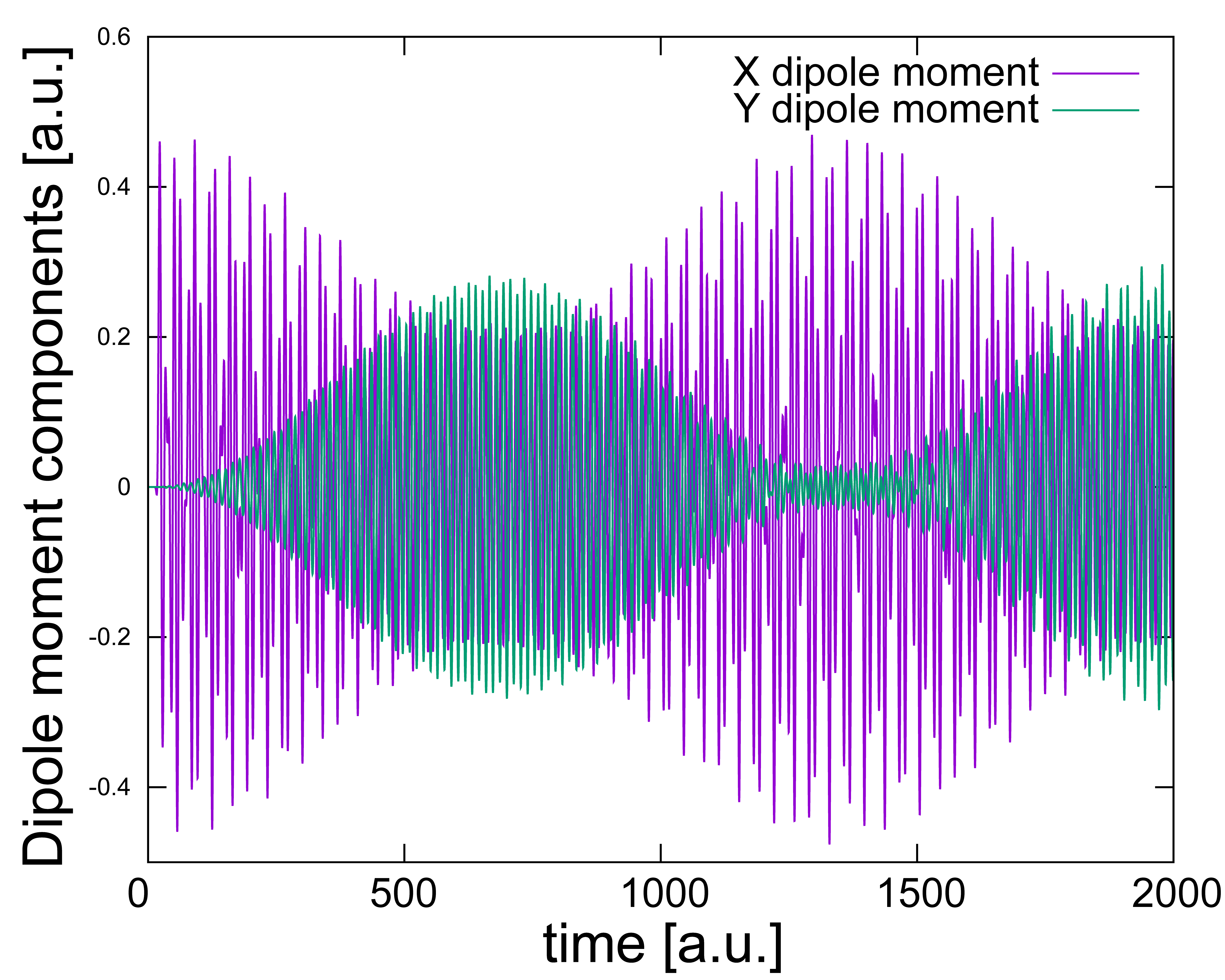}\\
    \includegraphics[width=.45\textwidth]{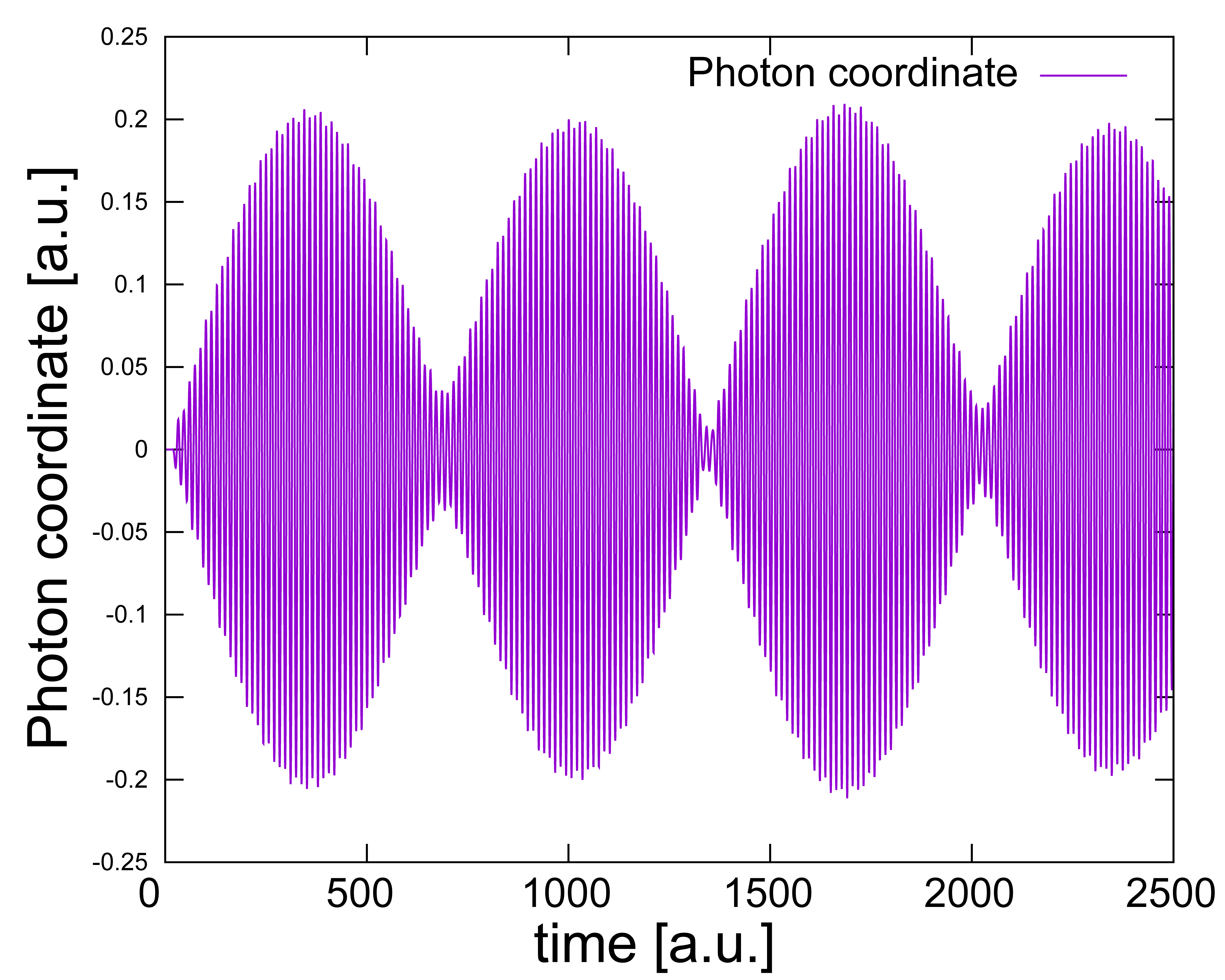}
    \caption{Dipole moment ${\bm{d}}$ components (upper panel) and photon coordinate $\frac{\braket{b^\dagger + b}}{\sqrt{2\omega}}$ (lower panel) for the dynamics depicted in \autoref{fig:(H2)2 ET}.
    The y component of the dipole moment provides a simple measure of the excitation transfer between the $\text H_2$ molecules, while the photon coordinate illustrates the photon excitation inside the cavity.}
    \label{fig:(H2)2 50AA, 0.76}
\end{figure}
\begin{figure}
    \centering
    \includegraphics[width=.45\textwidth]{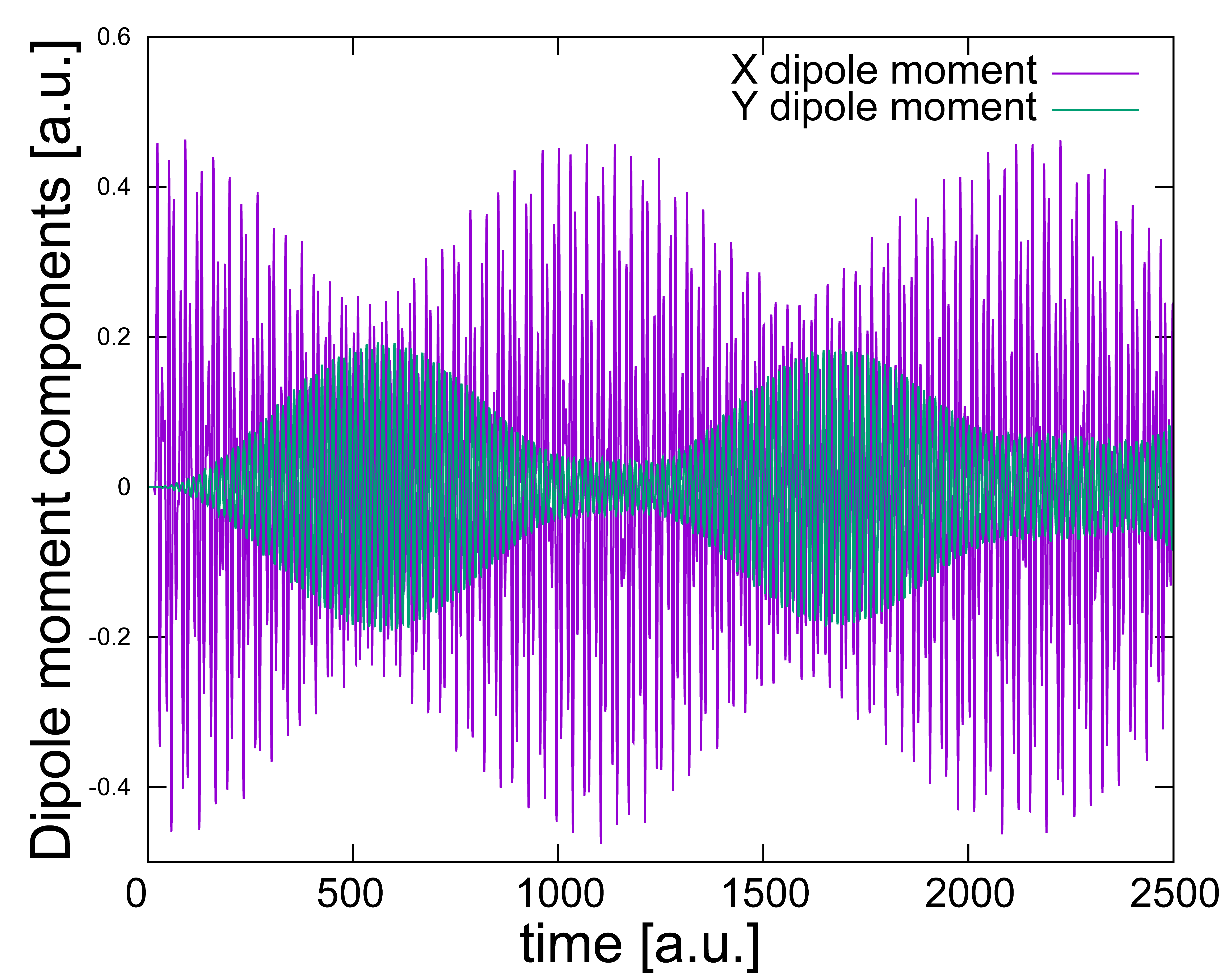}\\
    \includegraphics[width=.45\textwidth]{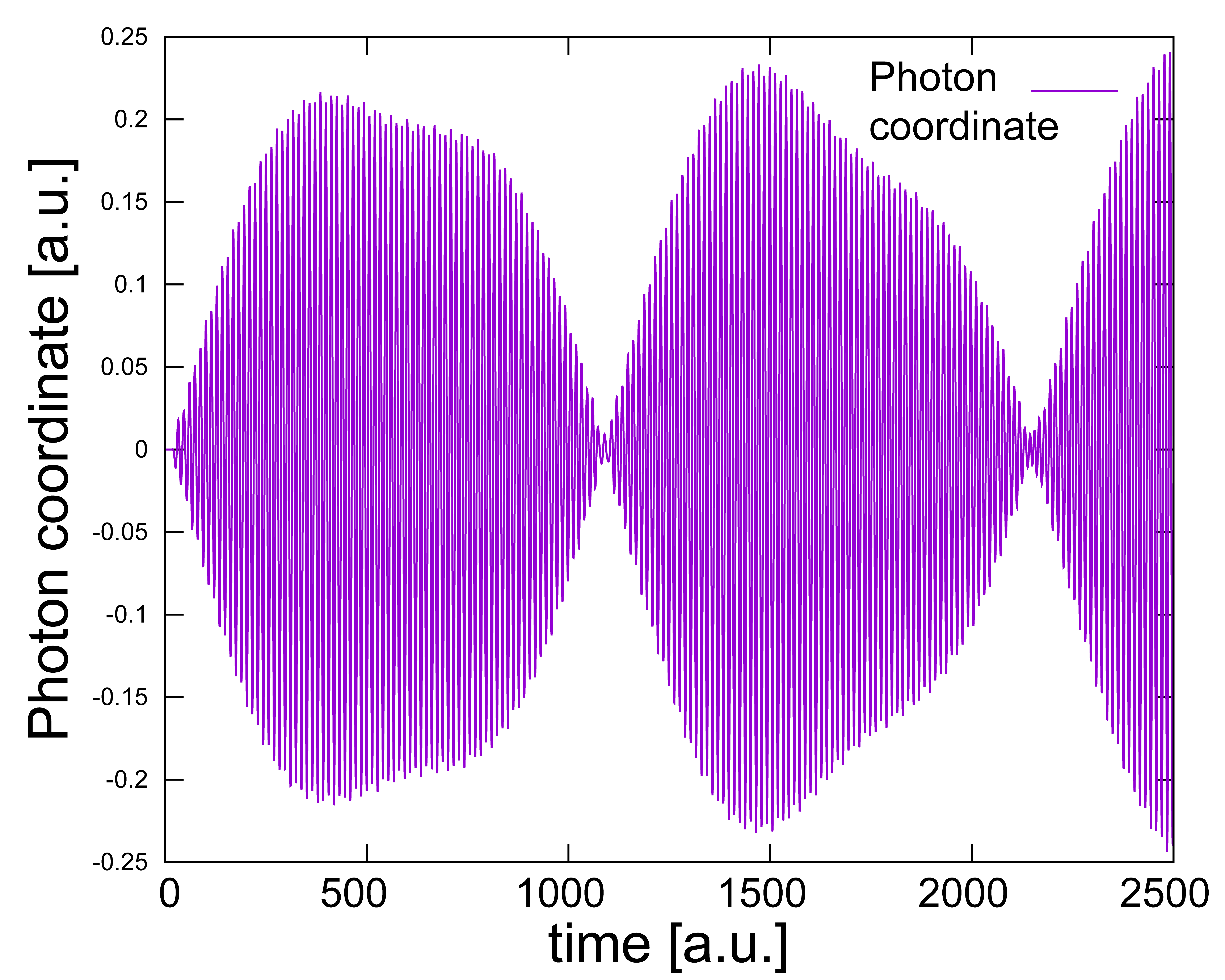}
    \caption{Dipole moment ${\bm d}$ components (upper panel) and photon coordinate $\frac{\braket{b^\dagger + b}}{\sqrt{2\omega}}$ (lower panel) for the dynamics of two slightly different $\text H_2$ molecules (of bond lengths \qtylist{0.76;0.78}{\angstrom}), with orientation, cavity field, and external pulse as in \autoref{fig:(H2)2 ET}. 
    The system is thus endowed with three distinct polaritonic states (lower, middle, and upper polaritons).
    Since the two undressed $\text H_2$ excitations are slightly different, the maxima in the photon coordinate envelope (second panel) are now different, contrary to \autoref{fig:(H2)2 50AA, 0.76}.}
    \label{fig:(H2)2 50AA, 0.76-0.78}
\end{figure}
It is interesting to notice that photon and matter observables show out-of-phase oscillations and seem to display a different time period, with \textit{two} maxima for $\braket{q}$ in a \textit{single} (back and forth) energy transfer between the $\text H_2$ molecules.
The reason for this (apparent) period difference is that each $\braket{q}$ maximum refers to \textit{one of the} $\text H_2$ transferring their excitation to the photon field.
This is illustrated more clearly using two slightly different molecules, stretching the bond length of the $\text H_2$ along y.
The dipole moment components and the photon coordinate for this system are illustrated in \autoref{fig:(H2)2 50AA, 0.76-0.78}.
The photon and matter degrees of freedom now exhibit the same characteristic time, but there are still two maxima in the photon coordinate for the back-and-forth energy transfer. 
However, the two peaks are now different as they refer to molecules \textit{with slightly different excitations} transferring energy to the optical field.
From \autoref{fig:(H2)2 50AA, 0.76-0.78}, it is also clear that there are more time scales involved compared to \autoref{fig:(H2)2 50AA, 0.76}, as can also be inferred from a simple Jaynes-Cummings (JC) analysis.
In the JC model, the system in \autoref{fig:(H2)2 ET} is described as two identical two-level oscillators resonantly coupled to the cavity field. 
There are then three different eigenstates: the upper $\ket{\text{UP}}$ and lower $\ket{\text{LP}}$ polaritons, and one dark state $\ket{\text{DS}}$ (of the same energy as the $\text H_2$ excited state and the photon)
\begin{align}
    \ket{\text{UP}} &= \frac{1}{\sqrt{2}}\bigg(\frac{\ket{e_1g_20}+\ket{g_1e_20}}{\sqrt{2}}+\ket{g_1g_21}\bigg)\\
    \ket{\text{DS}} &= \frac{\ket{e_1g_20}-\ket{g_1e_20}}{\sqrt{2}}\\
    \ket{\text{LP}} &= \frac{1}{\sqrt{2}}\bigg(\frac{\ket{e_1g_20}+\ket{g_1e_20}}{\sqrt{2}}-\ket{g_1g_11}\bigg),
\end{align}
where $e_p$ ($g_p$) refers to the p-th $\text H_2$ molecule in the excited (ground) state, and 0 (1) to the zero- (one-) photon states.
Notice that the dark state $\ket{\text{DS}}$, while showing no contribution from the photon field, has a non-zero transition dipole moment due to the different orientation of the molecules
\begin{equation}
    \braket{g_1g_20|\bm{d}|DS} = \big(\braket{g_1|x_1|e_1},-\braket{g_2|y_2|e_2},0\big)^T.
\end{equation}
After the pulse has passed, the system is still mainly in the ground state (linear excitation regime) superimposed with the polaritonic {and} dark states
\begin{equation}\label{eq:single molecule JC excitation}
    \ket{e_1g_20} = \frac{1}{2}\big(\ket{\text{UP}}+\ket{\text{LP}}+\sqrt{2}\ket{\text{DS}}\big).
\end{equation}
The interference between the involved states thus generates the modulation of the oscillation amplitude (quantum beats) in \autoref{fig:(H2)2 50AA, 0.76}.
There are indeed three fast time scales, associated with the excitation energies of the polaritonic and dark states, but only {two} slow time scales given by the Rabi splitting (UP-LP energy splitting) and the LP-DS energy difference.
In fact, the LP-DS and UP-DS energy difference is the same, and it is exactly half the Rabi splitting in the JC model.
On the other hand, when the hydrogens are different, as in \autoref{fig:(H2)2 50AA, 0.76-0.78}, three \textit{polaritonic} states are formed: the upper (UP), middle (MP), and lower (LP) polaritons.
In this case, three distinct fast time scales are present (from the MP-LP, MP-UP, and LP-UP energy differences), which explains why the photon revival in \autoref{fig:(H2)2 50AA, 0.76-0.78} differs from the first oscillation.
These simulations also demonstrate how the dark states can actively participate in the electron-photon dynamics of the system in the strong coupling regime.

The cavity photons thus provide a novel pathway for intermolecular energy transfer.
Electronic energy transfer in chemistry usually arises from a dipole-dipole (F{\"o}rster) or electronic-exchange (Dexter) mechanism and both are highly dependent on the distance and orientation of the involved molecules.\cite{scholes2003long}
Since the CC parametrization includes electron correlation, our method can describe intermolecular forces and is thus also suited for interacting molecules.
In \autoref{fig:(H2)2 5AA, 0.76}, we report the energy transfer between two $\text H_2$ molecules with the same parameters as in \autoref{fig:(H2)2 ET}, but for a shorter intermolecular distance $D=$ \qty[mode = text]{5}{\angstrom}.
The last panel of \autoref{fig:(H2)2 5AA, 0.76} shows the out-of-cavity (no QED) simulation of the x and y dipole moments, which illustrates the energy transfer due to the electronic coupling.
Inside the cavity (first two panels), both the electronic and polaritonic mechanisms are clearly involved, as we see the faster sinusoidal polaritonic energy transfer together with the electronic energy transfer.
Notice that the photon-mediated energy transfer is faster and more efficient than the electronic mechanism.
In \autoref{fig:(H2)2 5AA, 0.76-0.78}, we show the same simulation for two slightly different $\text H_2$ molecules.
Due to the different excitation energies, almost no energy transfer occurs without the photon field involvement (last panel), while two distinct photon maxima are highlighted due to the coupling of the different $\text H_2$ molecules to the optical device, as in \autoref{fig:(H2)2 50AA, 0.76-0.78}.
Polaritons thus provide an alternative energy-transfer channel subject to different requirements than standard electronic transfer.
Dexter energy transfer requires strong electronic couplings and is thus relevant only for very short distances, while the F{\"o}rster mechanism decays as the sixth power of the intermolecular separation $1/R^6$.
The  F{\"o}rster energy transfer also requires dipole-allowed molecular transitions and depends on the relative molecular orientation.
In addition, the involved excitations need to have similar energies: the fluorescence spectrum of the energy-donor molecule must have a significant spectral overlap with the absorption spectrum of the acceptor molecule.
On the other hand, the polaritonic mechanism is virtually distance-independent, only requiring bright (dipole-allowed) molecular excitations coupled to a quasi-resonant optical device.
In this case, the relative molecular orientation is less critical, while both excitations must hybridize by coupling to the same optical mode.
\begin{figure}
    \centering
    \includegraphics[width=.45\textwidth]{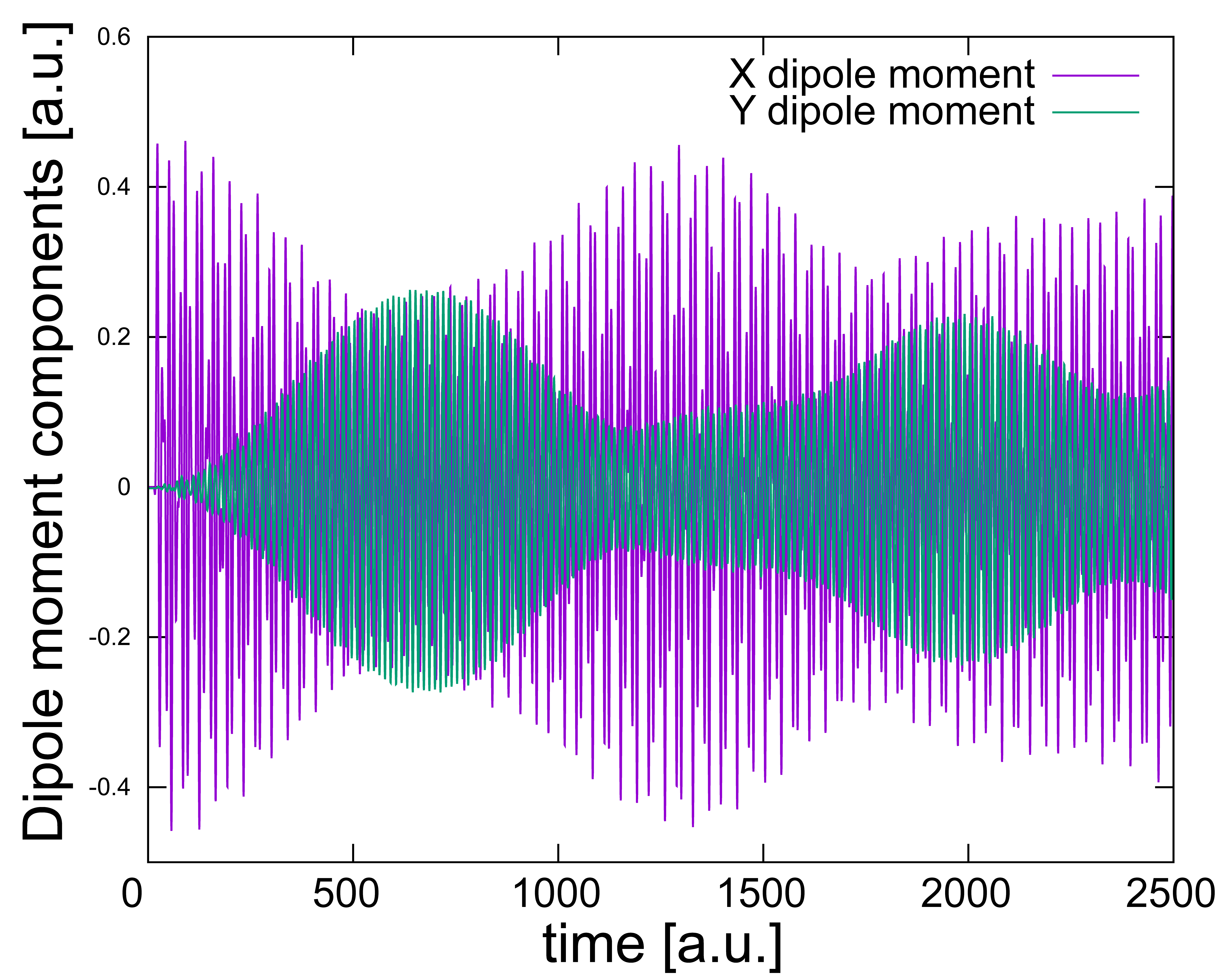}\\
    \includegraphics[width=.45\textwidth]{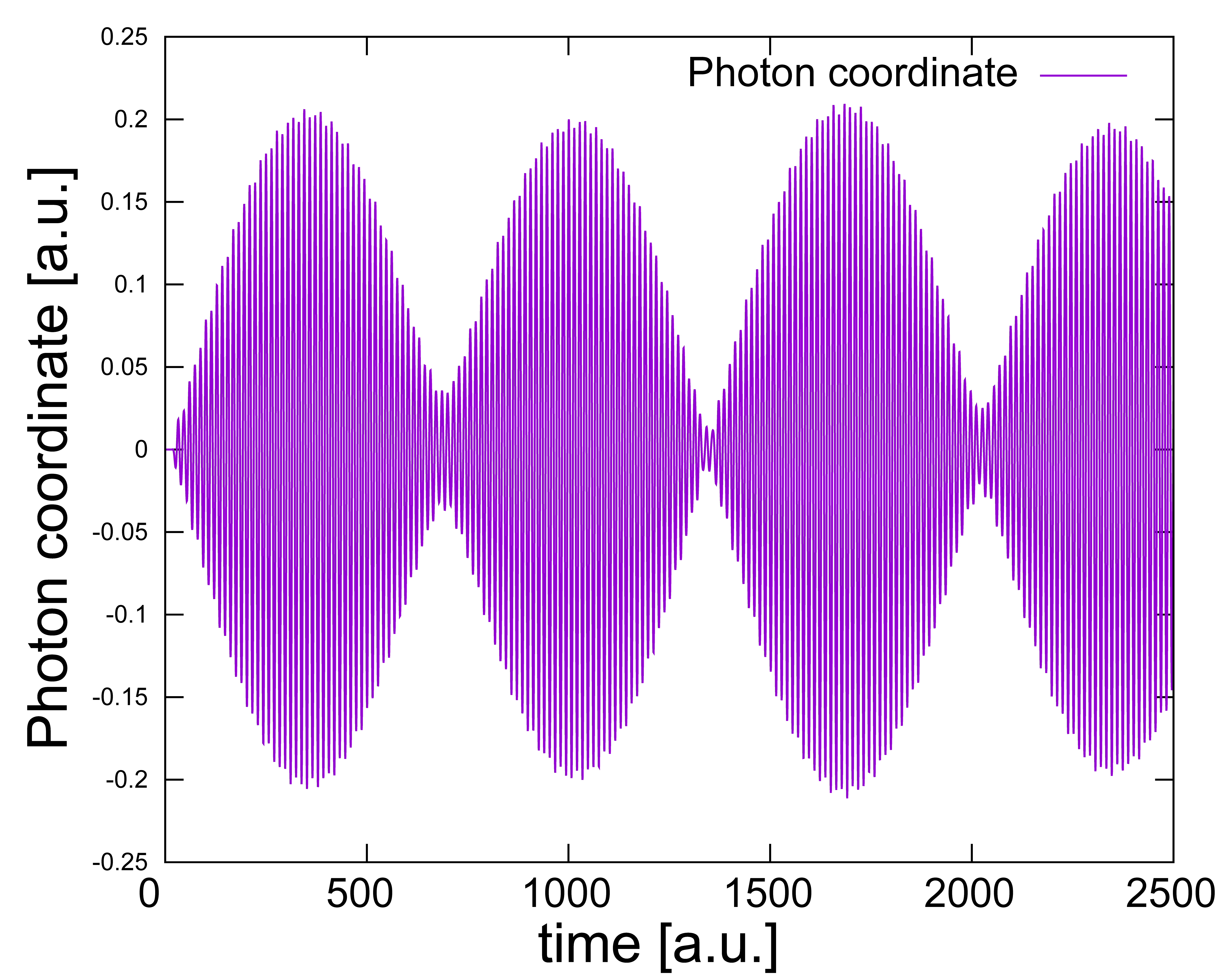}\\
    \includegraphics[width=.45\textwidth]{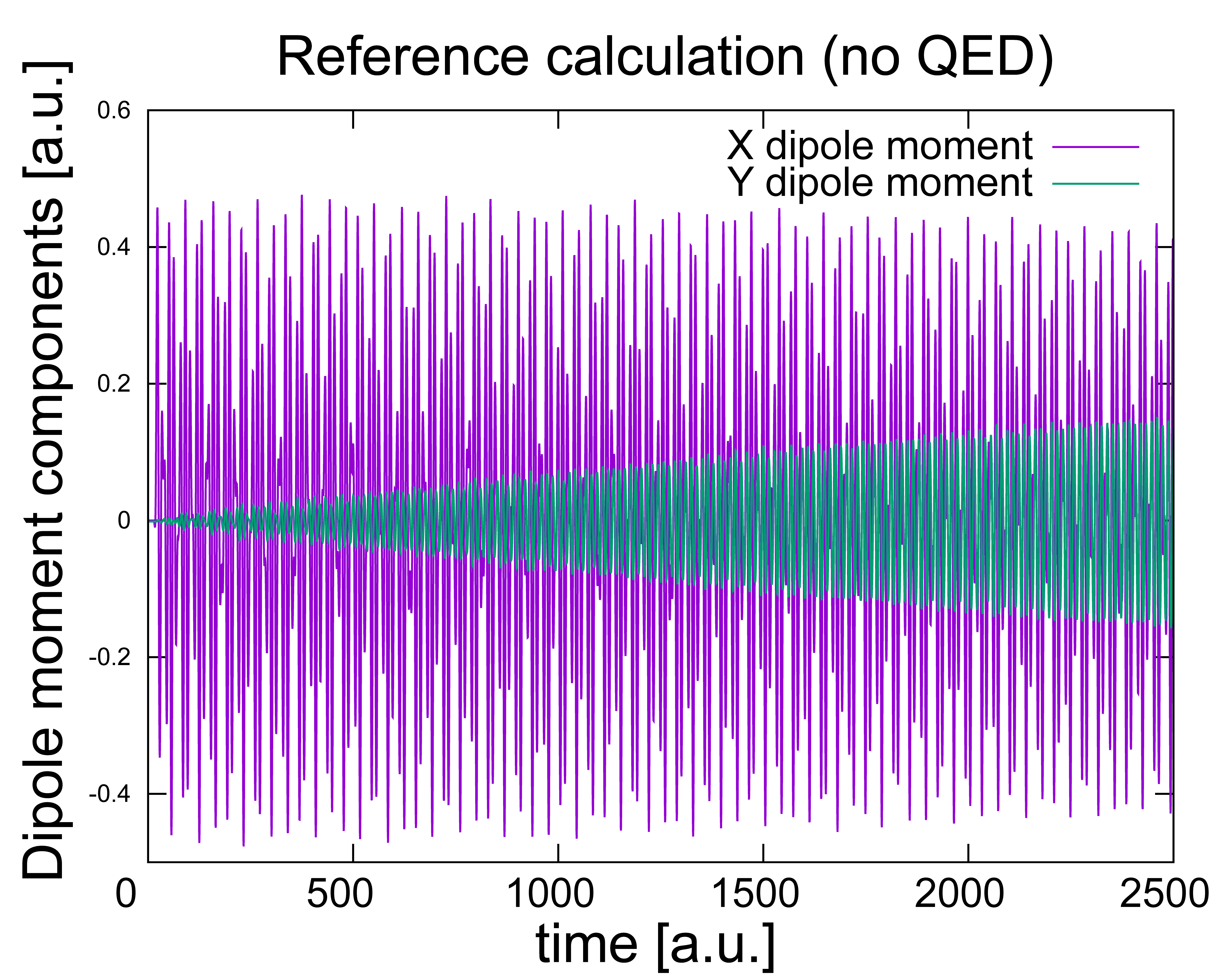}
    \caption{Dipole moment ${\bm d}$ components (upper panel), photon coordinate $\frac{\braket{b^\dagger + b}}{\sqrt{2\omega}}$ (middle panel), and reference dipole moment simulation (no QED, lower panel) for the dynamics of two perpendicular but identical $\text H_2$ molecules at a distance of \qty[mode = text]{5}{\angstrom}. 
    Due to electronic coupling, there is an energy transfer even in the absence of the photon field mediation.}
    \label{fig:(H2)2 5AA, 0.76}
\end{figure}
\begin{figure}
    \centering
    \includegraphics[width=.43\textwidth]{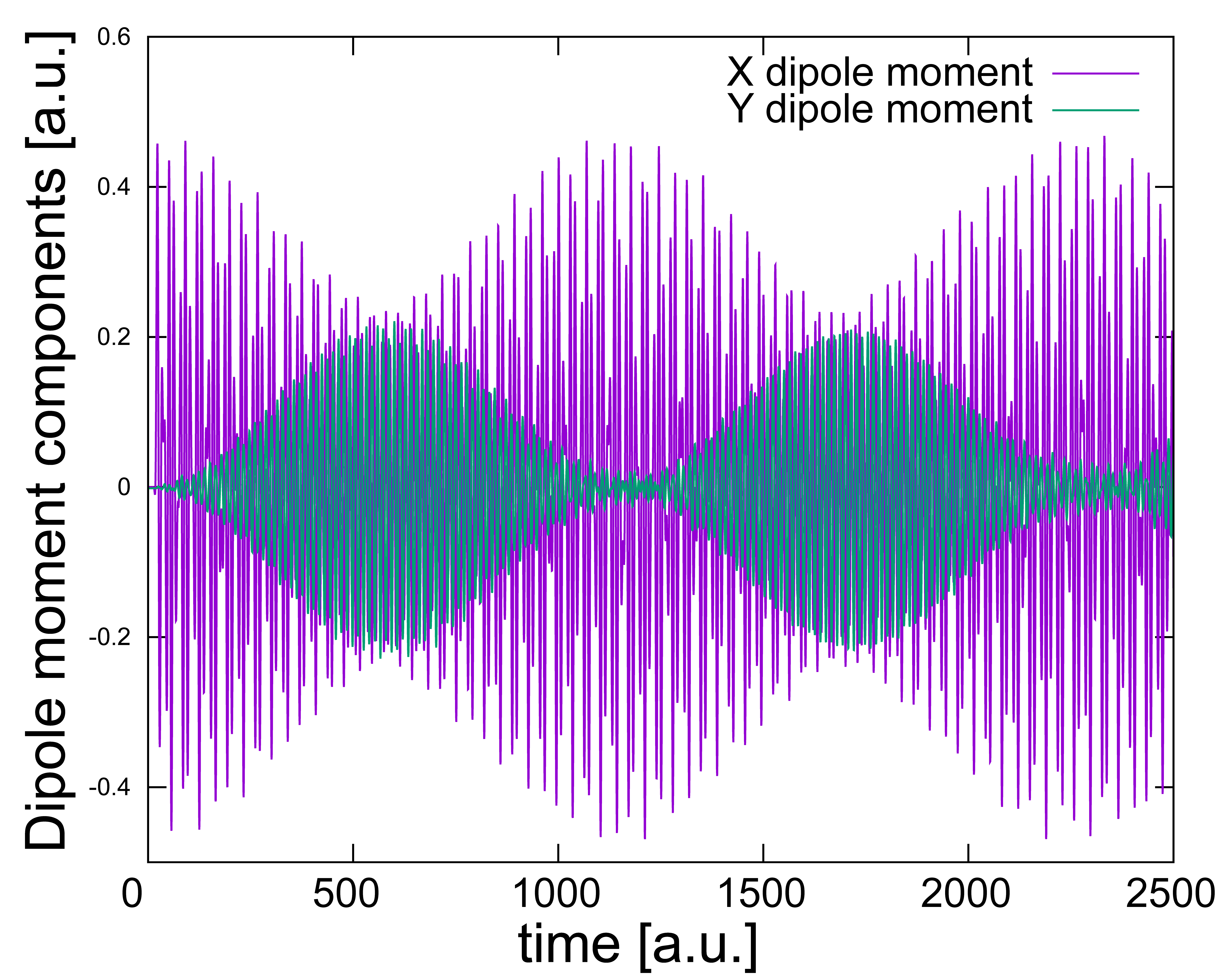}\\
    \includegraphics[width=.43\textwidth]{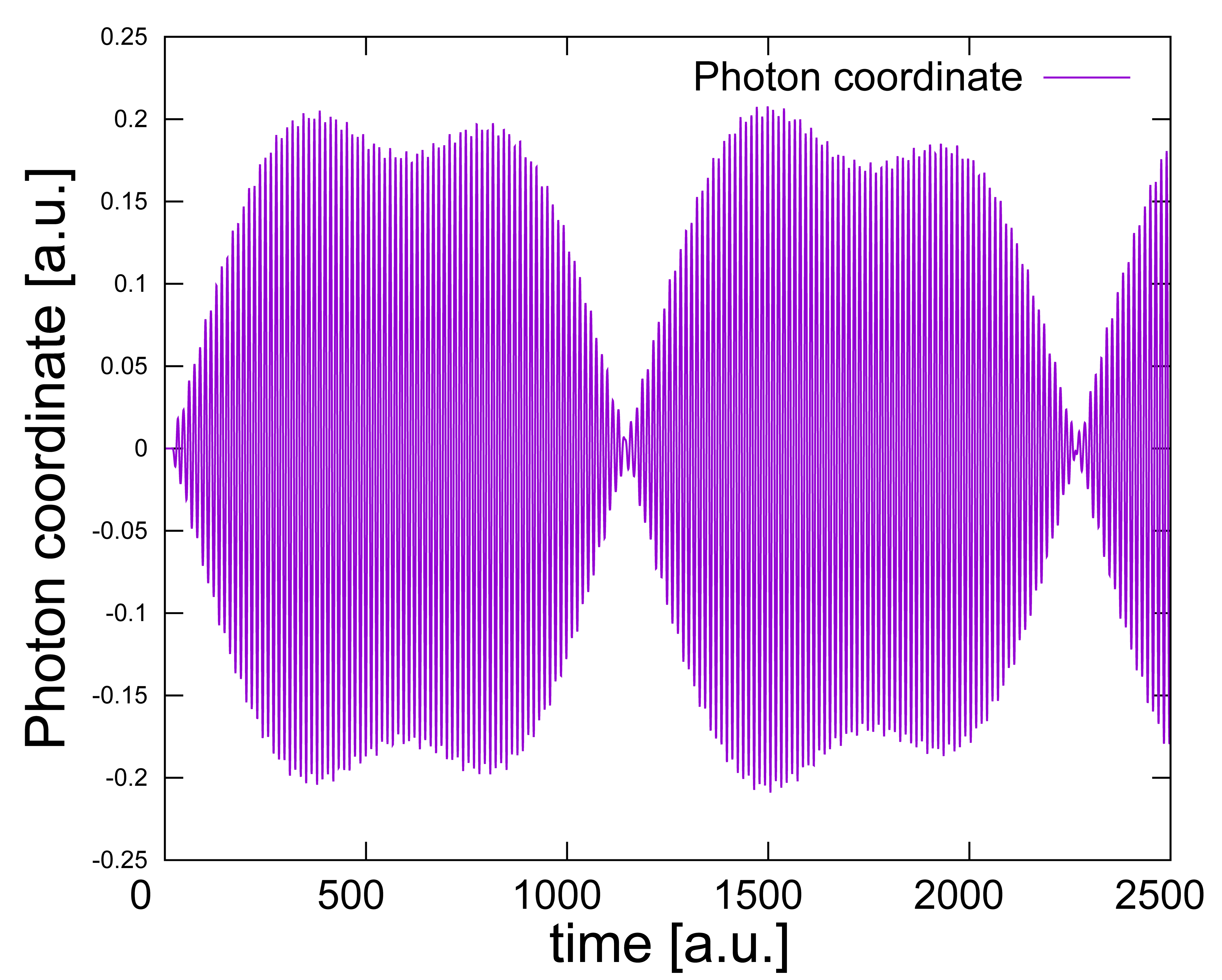}\\
    \includegraphics[width=.43\textwidth]{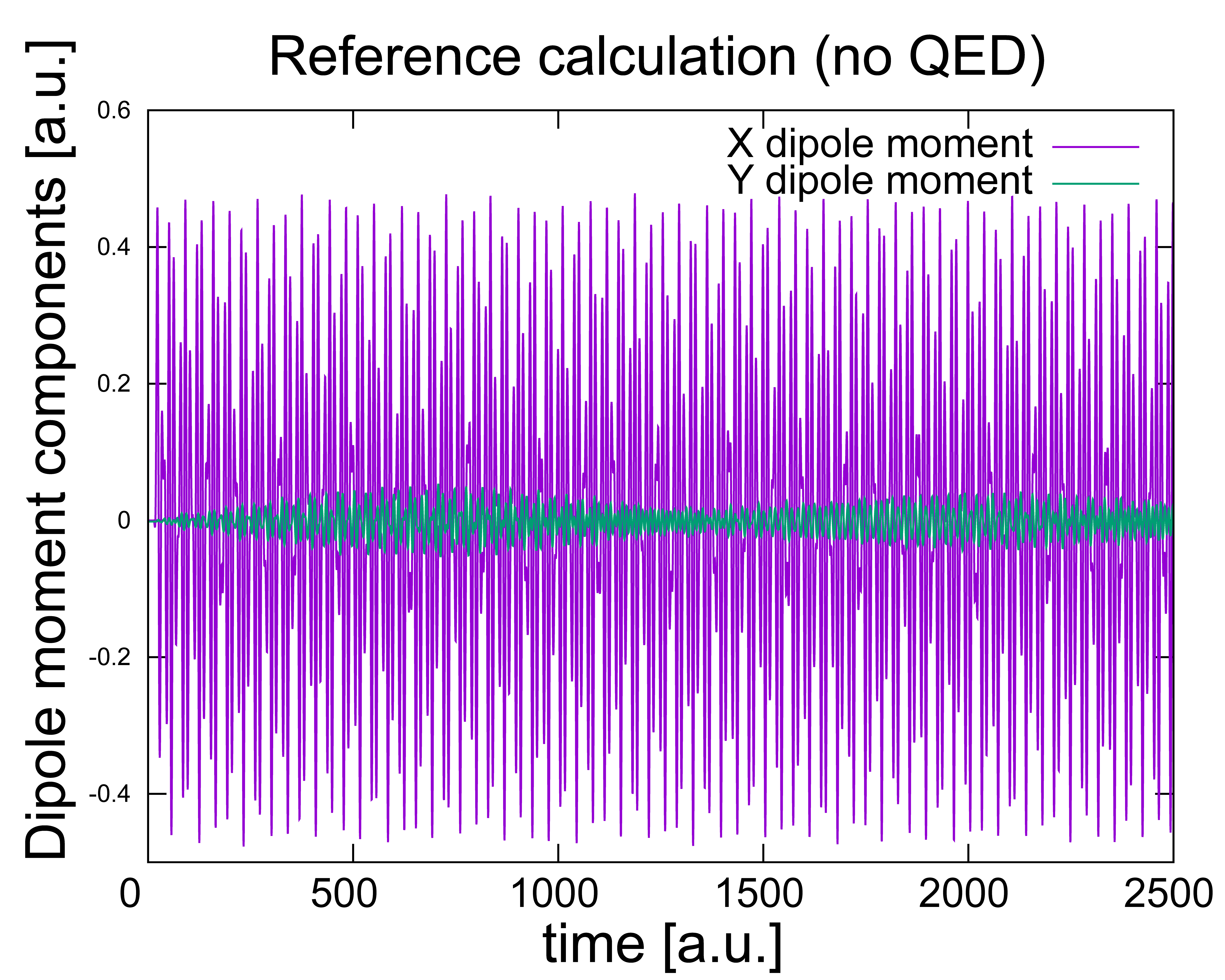}
    \caption{Dipole moment $\bm d$ components (upper panel), photon coordinate $\frac{\braket{b^\dagger + b}}{\sqrt{2\omega}}$ (middle panel), and reference dipole moment simulation (no QED, lower panel) for the dynamics of two perpendicular $\text H_2$ molecules with a slightly different bond length at a distance of \qty[mode = text]{5}{\angstrom}. 
    The electronic coupling is highly dependent on the excitation energy, the orientation, and the distance between the molecules and, compared to \autoref{fig:(H2)2 5AA, 0.76}, little energy exchange occurs without the photon mediation.}
    \label{fig:(H2)2 5AA, 0.76-0.78}
\end{figure}

Finally, we investigated how the electron-photon dynamics changes with the light-matter coupling strength $\lambda$ and the number of identical replicas $N$ of the system (see also the Supporting Information). 
Since the time scale of the photon-mediated energy transfer depends on the Rabi splitting, it is inversely proportional to $\lambda$ and the square root of $N$.
That is, the time scale depends only on the inverse of the \textit{collective} coupling strength $\lambda \sqrt{N}$, which means that such processes are relevant in the thermodynamic limit $N\to\infty$, $\lambda\sqrt{N}=\text{const.}$ and the time scales can be reduced by simply increasing the molecular concentration in the optical device.
This is shown in the Supporting Information for the systems of \autoref{fig:(H2)2 50AA, 0.76} and \ref{fig:(H2)2 50AA, 0.76-0.78} (identical and different $\text H_2$ molecules).
At the same time, the energy absorbed and the dipole moment oscillation amplitudes scale linearly when increasing the number of replicas $N$ subject to \textit{the same} {external} pulse.
This is physically reasonable, and it is correctly modeled via the size intensivity and extensivity of the CC parametrization for the electrons.\cite{helgaker2013molecular, skeidsvoll2023comparing}
On the other hand, the oscillation amplitudes of the photon coordinate scales as $\sqrt{N}$.
The amplitude of matter and photon observables thus scale differently with $N$, while both are fundamentally independent on $\lambda$.
These results, therefore, suggest that a relevant amount of energy is stored in the increasing number of dark states.

\subsection{Photon generation and modified electron-photon dynamics \textit{via} classical pulse sequences}

\begin{figure}
    \centering
    \includegraphics[width=.45\textwidth]{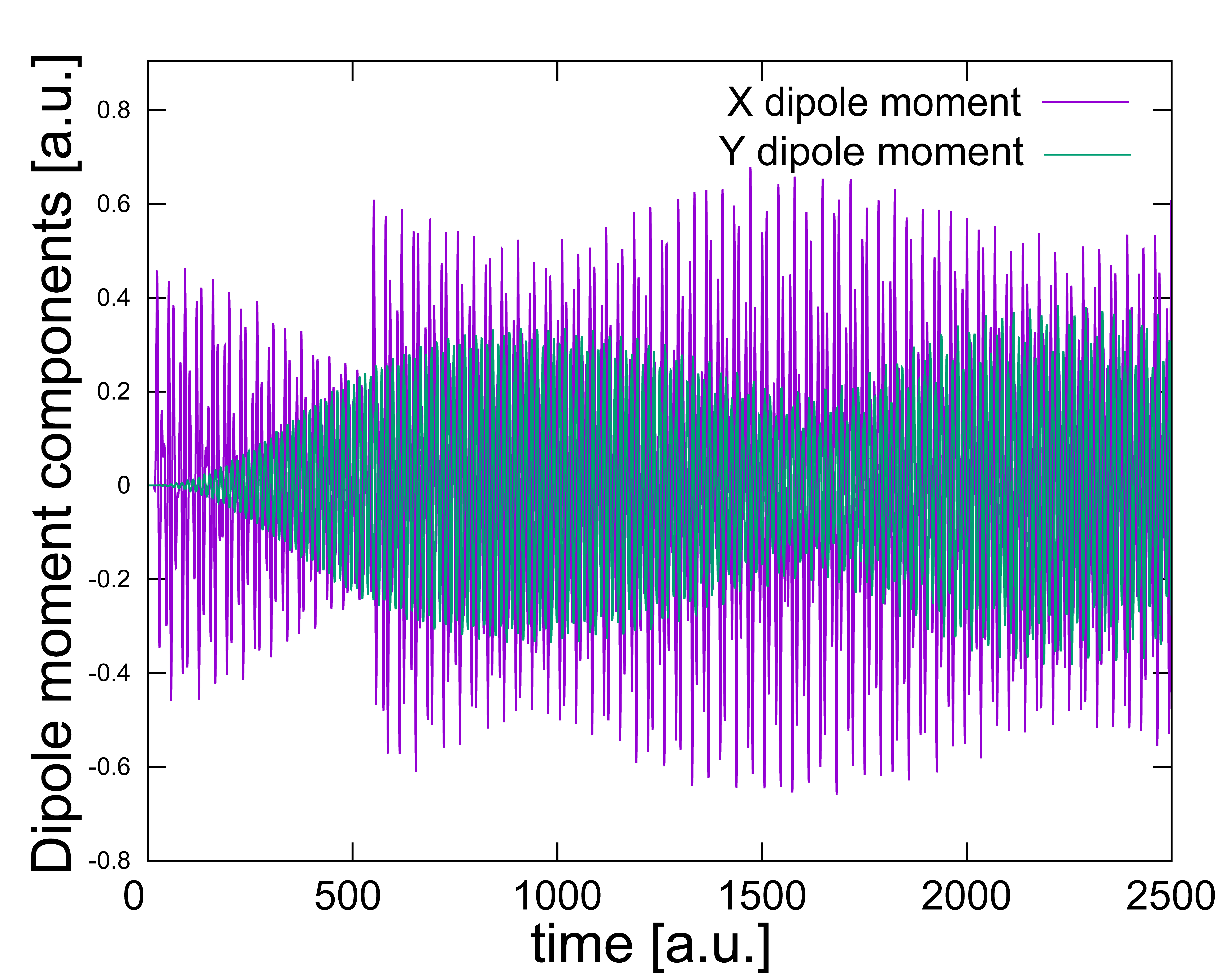} \\ 
    \includegraphics[width=.45\textwidth]{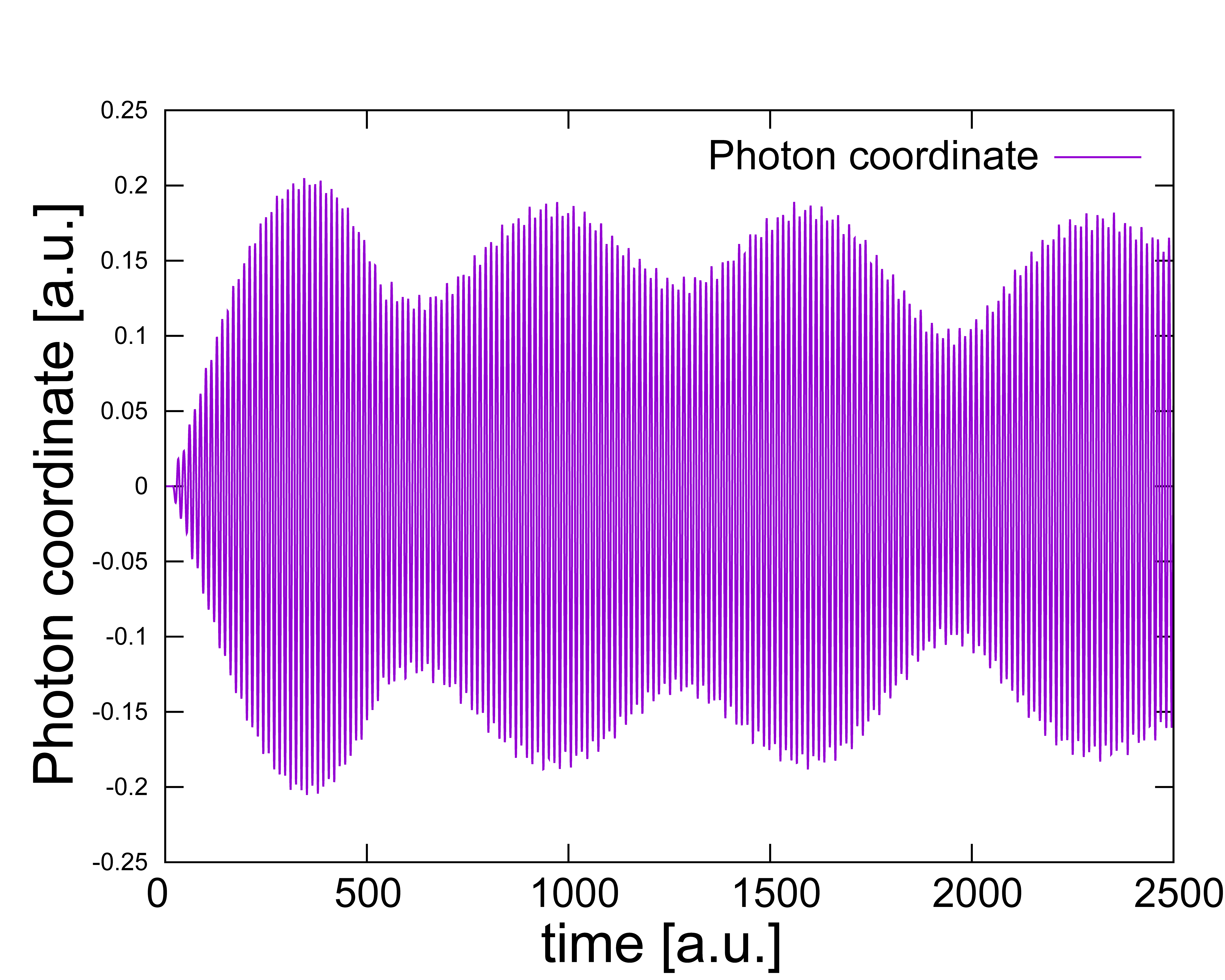} \\
    \includegraphics[width=.45\textwidth]{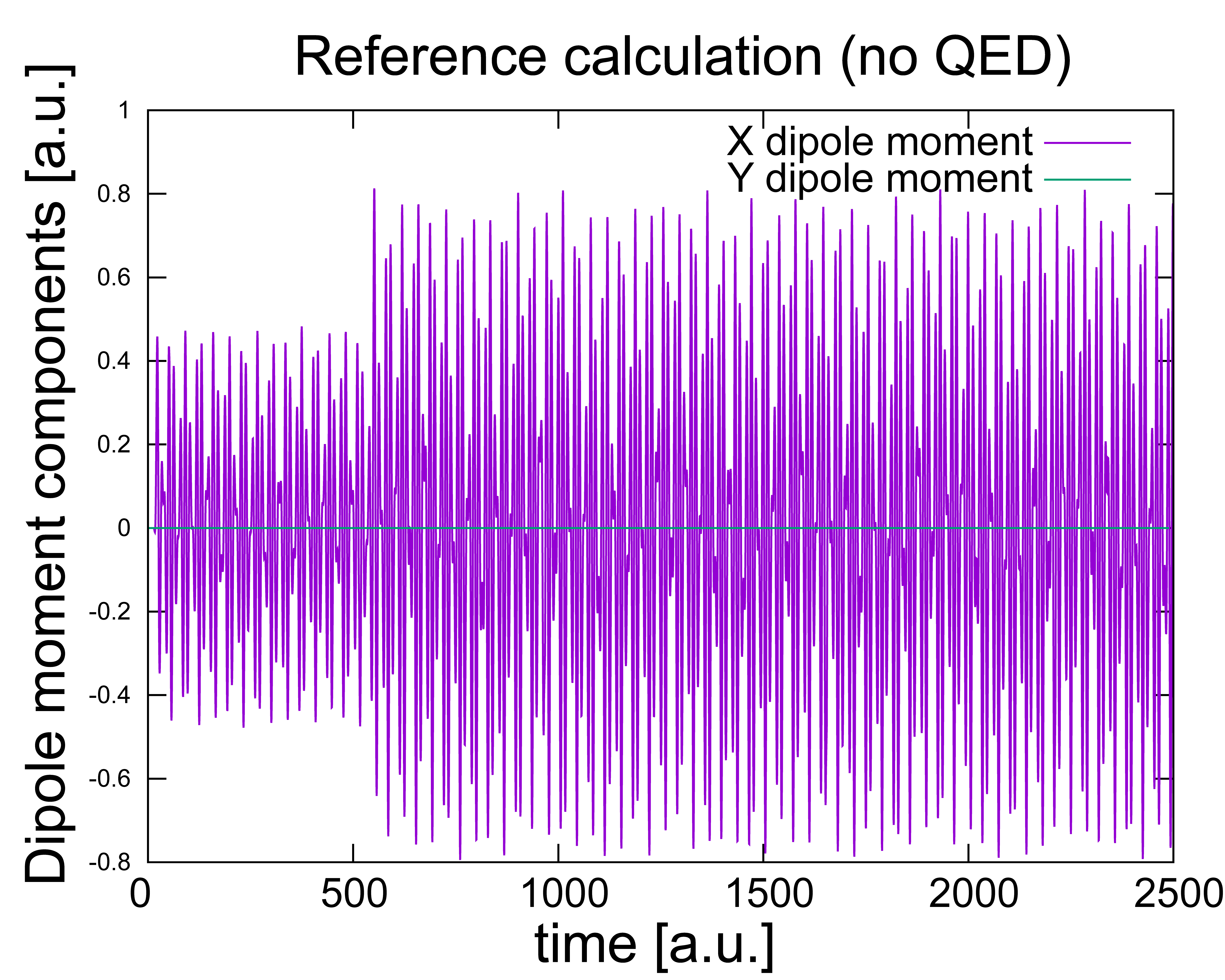}
    \caption{Dipole moment $\bm d$ components (upper panel), photon coordinate $\frac{\braket{b^\dagger + b}}{\sqrt{2\omega}}$ (middle panel), and reference dipole moment simulation (no QED, lower panel) for the dynamics of two perpendicular but identical $\text H_2$ molecules at a distance of \qty[mode = text]{50}{\angstrom}. 
    The system is excited via two ultrashort pulses centered at $t = $ \qtylist{20;550}{\atomicunit}, where the second pulse is switched on just before the maximal energy transfer between the two $\text H_2$ molecules shown in \autoref{fig:(H2)2 50AA, 0.76}.
    }
    \label{fig: 2 pulses }
\end{figure}

We now focus on the energy transfer and electron-photon dynamics induced by multiple classical electric pulses.
In \autoref{fig: 2 pulses }, we report the time evolution of the dipole moment and photon coordinate induced by two short classical electric pulses centered at \qtylist{20;550}{\atomicunit} for the same system depicted in \autoref{fig:(H2)2 ET}.
The second pulse is switched on just before the maximum transfer between the two $\text H_2$ molecules, and the undressed (out-of-cavity) electronic dynamics is reported in the last panel for reference.
Notice that no energy transfer occurs outside the cavity because of the large separation between the molecules, as can be inferred from the absence of the y component of the dipole moment in the lowest panel.
The $\text H_2$ oriented along the external electric field (x-axis) is excited again by the second pulse, causing a sharp change of the x component of the dipole moment.
At the same time, the photon coordinate varies smoothly, but its envelope is modified compared to \autoref{fig:(H2)2 50AA, 0.76}.
Without the second pulse, the photon coordinate amplitude would decrease almost to zero, as shown in \autoref{fig:(H2)2 50AA, 0.76}. 
After the $\text H_2$ along x is excited by the second pulse, the photon coordinate increases again, and its envelope amplitude never decreases to zero.
This means that, through classical electric fields and their coupling \textit{only to matter}, we generated photons in the optical device that the molecules will not completely absorb as in the vacuum Rabi oscillations of \autoref{fig:(H2)2 50AA, 0.76}.

\begin{figure}
    \centering
    \includegraphics[width=.45\textwidth]{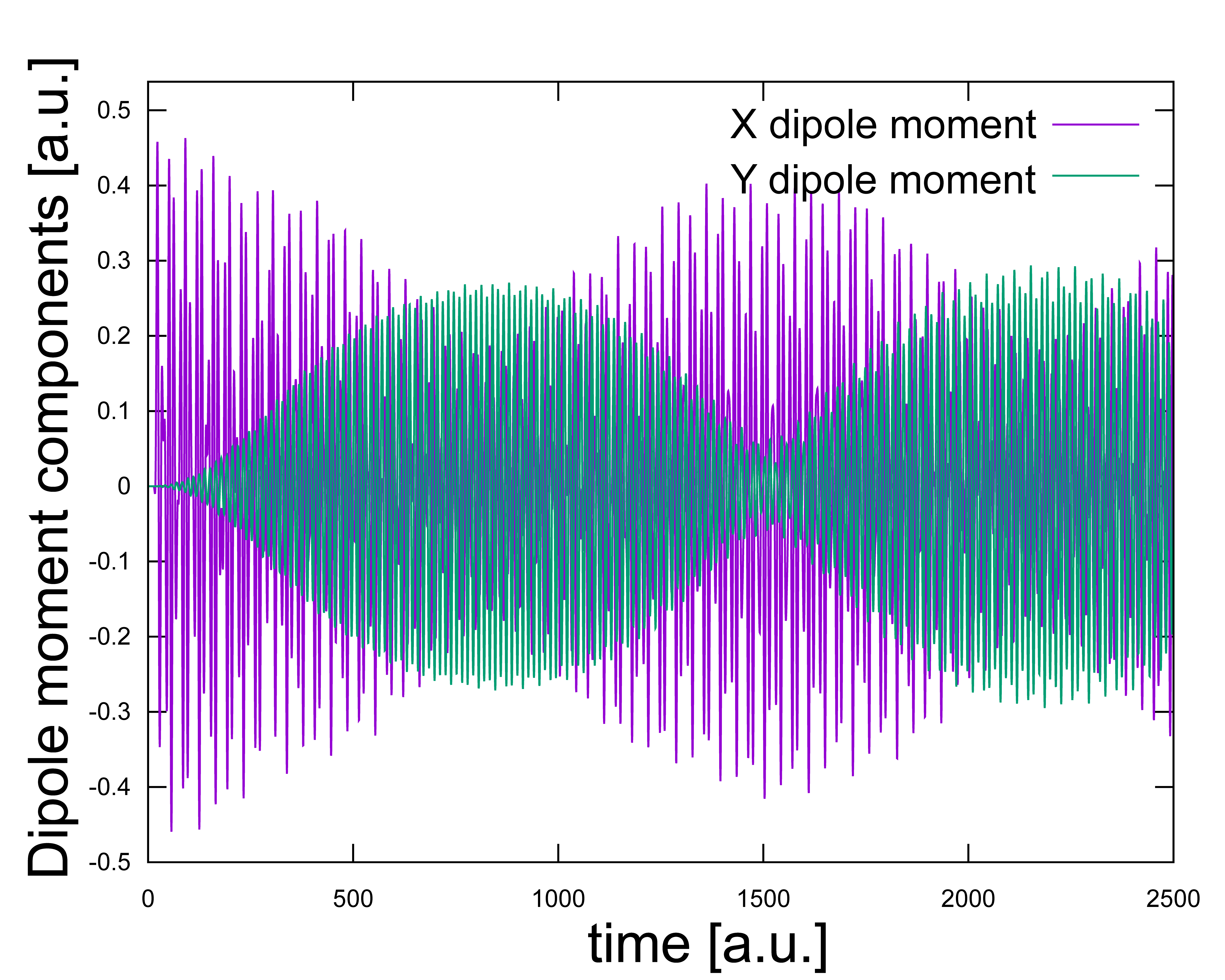}\\ 
    \includegraphics[width=.45\textwidth]{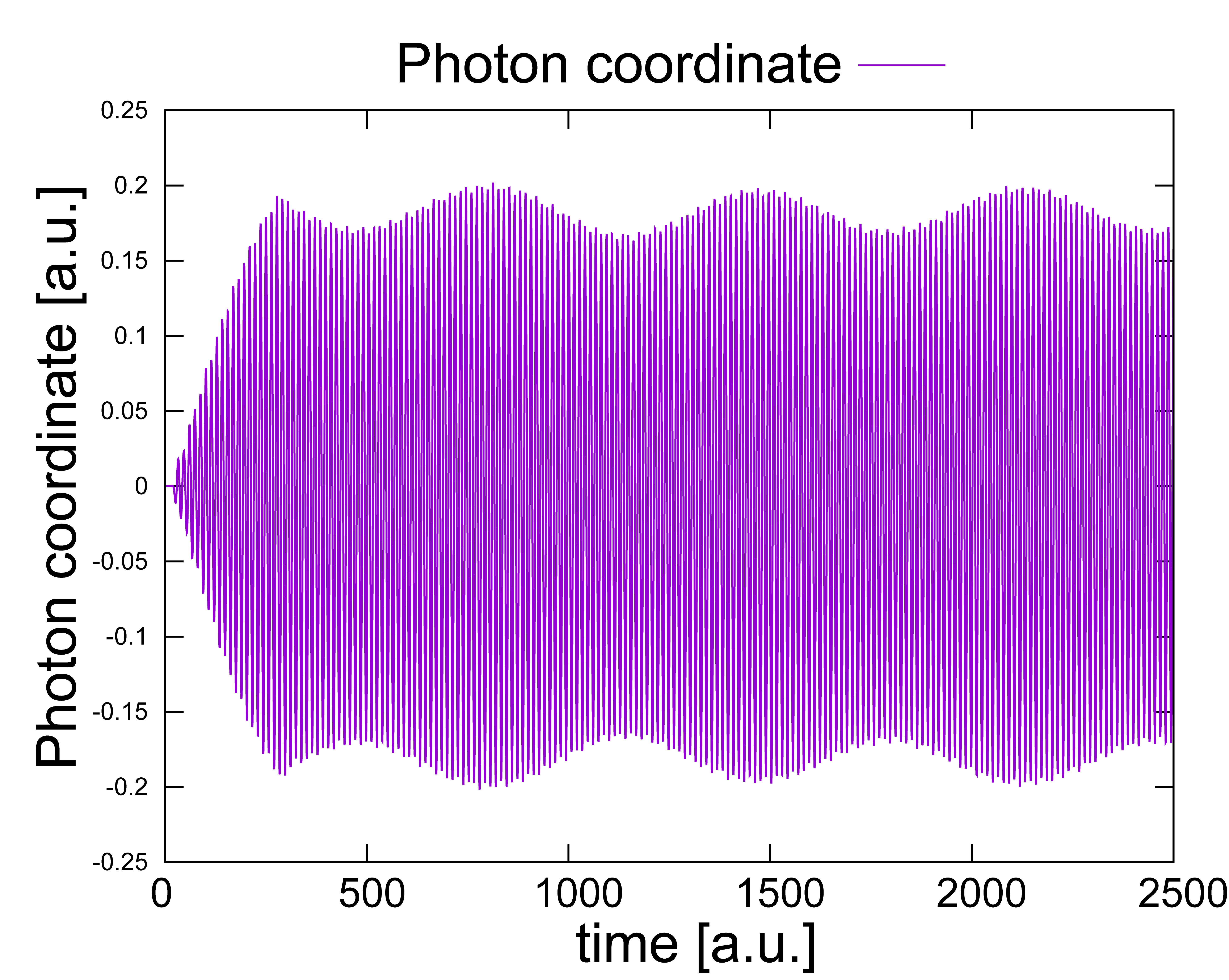}\\
    \includegraphics[width=.45\textwidth]{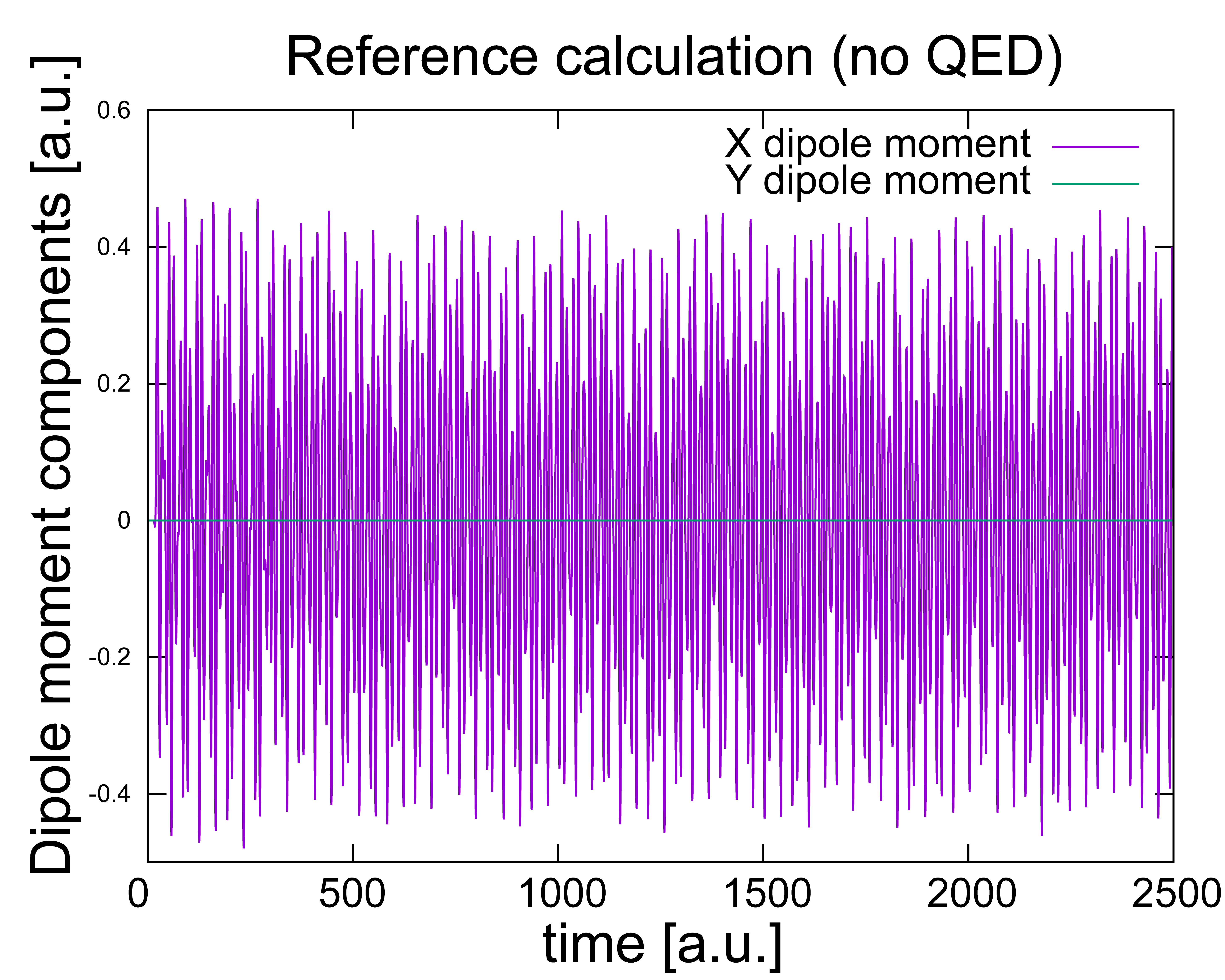}
    \caption{Dipole moment $\bm d$ components (upper panel), photon coordinate $\frac{\braket{b^\dagger + b}}{\sqrt{2\omega}}$ (middle panel), and reference dipole moment simulation (no QED, lower panel) for the dynamics of two perpendicular but identical $\text H_2$ molecules at a distance of \qty[mode = text]{50}{\angstrom}. 
    The system is excited via two ultrashort pulses centered at $t = $  \qtylist{20;300}{\atomicunit}, where the second pulse is switched on roughly at the maximum amplitude of the photon coordinate $q$ shown in \autoref{fig:(H2)2 50AA, 0.76}.
    }
    \label{fig: 2 pulses shif}
\end{figure}

Another interesting point of the dynamics in \autoref{fig:(H2)2 50AA, 0.76} is the photon coordinate maximum, which occurs before the energy transfer between the $\text H_2$ is complete.
In \autoref{fig: 2 pulses shif}, we show the dynamics of the system under the influence of two pulses centered at \qtylist{20;300}{\atomicunit}.
The central time of the second pulse now roughly corresponds to the first maximum of $\braket{q}$ in \autoref{fig:(H2)2 50AA, 0.76}.
In \autoref{tab:energy H2 2 pulses}, we report the energy of the system (inside and outside the cavity) before and after each pulse.
It is interesting that the optical device changes the interaction of the second pulse with the molecules.
In the reference (no QED) simulation, the second pulse triggers a spontaneous emission as the system has lost energy after the interaction. 
Nevertheless, its effect is small enough that no sharp change in the dipole moment is shown in the last panel of \autoref{fig: 2 pulses shif}.
\begin{table}[!ht]
    \centering
    \caption{Energy of the $(\text H_2)_2$ system before and after two ultrashort external electric pulses centered at $t =$ \qtylist{20;300}{\atomicunit}.
    For the undressed electrons, the molecule oriented along y does not participate in the simulation, and the second pulse stimulates energy emission. 
    In the QED simulation, after some energy has been transferred between the two $\text H_2$ molecules, the second pulse (slightly) excites the molecule along x again.}
    \begin{tabular}{|c|c|c|c|}
        \hline\text{Energy [a.u.]}&\text{Initial} &\text{Pulse 1} &\text{Pulse 2}\\
        \hline \text{RT-QED-CC} & -2.32991 & -2.31173 & -2.31083 \\
        \hline \text{RT-CC} & -2.32998 & -2.31180 & -2.31296 \\
        \hline
    \end{tabular}
    \label{tab:energy H2 2 pulses}
\end{table}
For the photon-dressed electrons, the molecules again gain energy from the pulse, as can be inferred from \autoref{tab:energy H2 2 pulses}.
In this case, the dipole moment along x still does not show a sudden envelope modification, but there is a sharp change in the envelope of the photon coordinate.
As in \autoref{fig: 2 pulses }, the amplitude of the photon coordinate oscillations does not reduce to zero, contrary to \autoref{fig:(H2)2 50AA, 0.76}, which means that there is always a nonzero number of photons inside the cavity.
However, the photon coordinate amplitude is larger than \autoref{fig: 2 pulses }, and the second $\text H_2$ molecule almost completely deexcites, contrary to the significant dipole moment along y shown in \autoref{fig: 2 pulses } after a complete oscillation.
Therefore, these results show that multiple classical pulses can trigger nontrivial photon dynamics in a quantum optic device and alter the photon-mediated energy transfer between molecules in the strong coupling regime.

\vfill

\subsection{Intramolecular electron-photon dynamics}
\begin{figure}
    \centering
    \includegraphics[width=.5\textwidth]{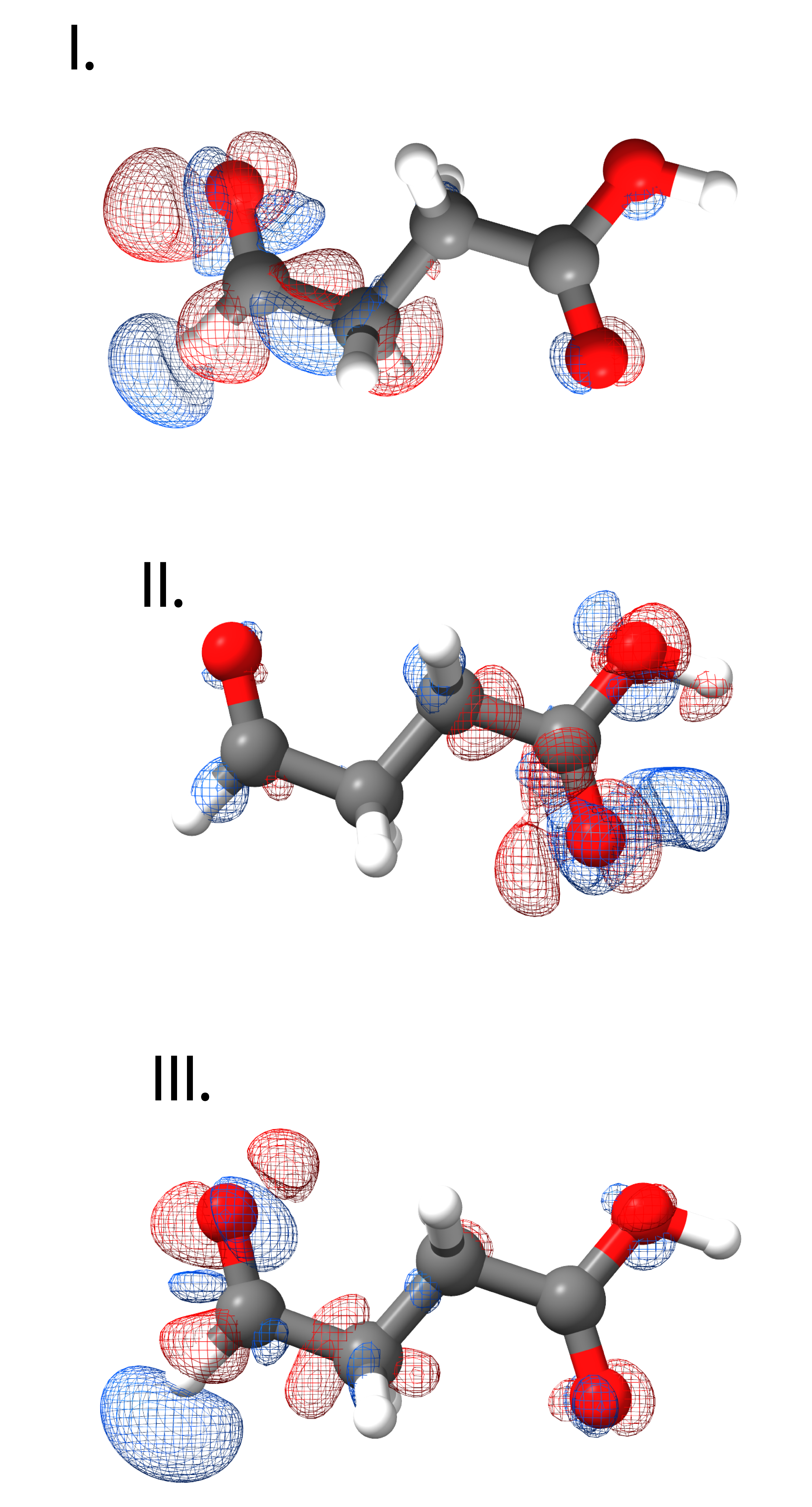}
    \caption{Ground to excited state transition densities for three selected excited states of the succinic semialdehyde $\text C_4\text H_6\text O_3$.
    The excitation energies and the transition moments are reported in \autoref{tab:states intramol}.
    The states I and III are mainly associated with the aldehyde moiety, while the state II is an excitation of the acidic group.}
    \label{fig:states intramol}
\end{figure}

\begin{figure}
    \centering
    \includegraphics[width=.45\textwidth]{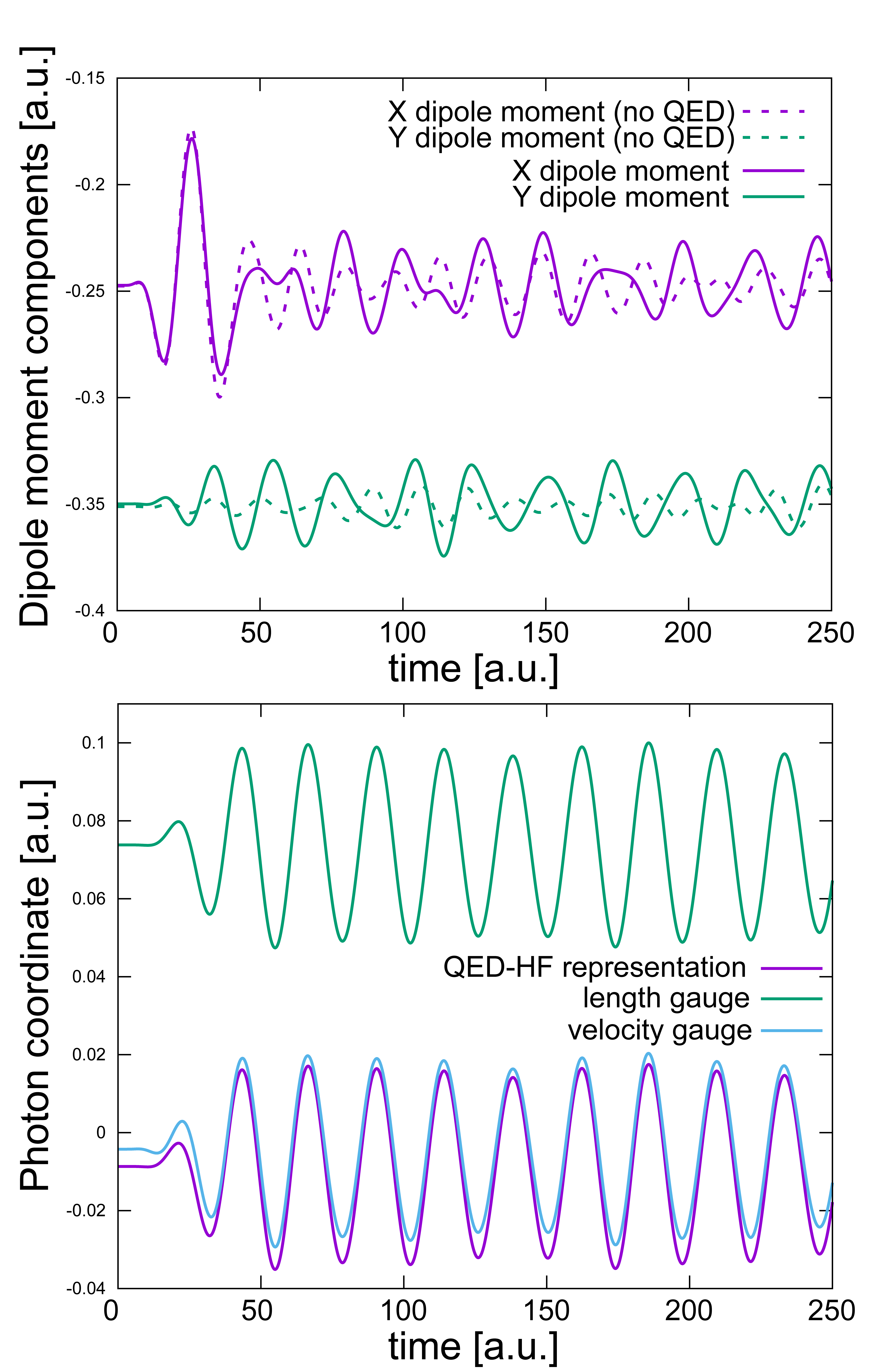}
    \caption{Time evolution of the x and y components of the dipole moment inside (solid lines) and outside (reference, dotted lines) the cavity for the succinic semialdehyde $\text C_4\text H_6\text O_3$.
    The cavity photon is resonant to the state II of \autoref{tab:states intramol}, and its polarization is along the xy axis, and thus coupled to all 3 states.
    The enhanced fluctuations (compared to the reference calculation) along y prove a larger cavity-induced energy transfer from state II to states II and III.}
    \label{fig:intramolecular energy dipole transfer}
\end{figure}

While we have focused so far on intermolecular energy transfer, the described effects apply equally to intramolecular processes.
As a chemically interesting {system}, we focus on the succinic semialdehyde $\text C_4\text H_6\text O_3$ illustrated in \autoref{fig:states intramol}.
In \autoref{fig:states intramol}, we also report the transition density of three selected electronic excited states, whose excitation energy and dipole moments are reported in \autoref{tab:states intramol} (computational details and additional data can be found in the Supporting Information).
\begin{table}[!ht]
    \centering
    \caption{Excitation energies and product of the CCSD left and right transition dipole moments along x, y, and z $\braket{L|d_p|CC}\braket{\Lambda|d_p|R}$ for the excited states of the succinic semialdehyde $\text C_4\text H_6\text O_3$ depicted in \autoref{fig:states intramol}.}
    \begin{tabular}{|c|c|c|c|c|}
        \hline state&energy &x&y&z  \\
        \hline I&7.144 &0.0035&0.1041&0.0002 \\
        \hline II&7.407 &0.3813&0.0066&0.0000  \\
        \hline III&7.802 &0.1489&0.1681&0.0010  \\
        \hline
    \end{tabular}
    \label{tab:states intramol}
\end{table}
The states I and III are primarily associated with the aldehyde moiety, while the state II is an excitation of the acidic group.
The molecule is then excited using a pulse centered at the energy of state II and polarized along x, thus coupling to the states II and III. 
For the QED simulation, the optical device is tuned to the undressed (out-of-cavity) excitation III reported in \autoref{tab:states intramol}, with moderate coupling strength $\lambda = $ \qty{0.05}{\atomicunit} and polarization along the xy-axis. 
The photon field is thus coupled to all 3 reported states, while the other excitations of the molecule are less relevant as they are highly detuned and carry lower transition dipoles.
The inspection of the density displacement after the interaction with the pulse shows that both the aldehyde and the acid group are excited.
The acidic moiety shows a more significant density displacement (and thus excitation), as expected since the pulse carrier frequency is tuned to state II and from its larger transition moment along x (see also the Supporting information).
In the first panel of \autoref{fig:intramolecular energy dipole transfer}, we report the time evolution of the x and y components of the dipole moments inside (solid lines) and outside (reference, dotted lines) the cavity.
The results clearly show larger fluctuations for the y dipole moment in the QED environment compared to the non-cavity case.
The states I and III are thus more involved in the polaritonic system than in the purely electronic dynamics.
In addition, the absorbed energies ($5.554 \times 10^{-5}$ a.u. and $4.640 \times 10^{-5}$ a.u. for the QED and reference electronic calculation, respectively) show a more favorable interaction of the external pulse with the polaritonic system.
Therefore, the restructuring of the energy landscapes promoted by the light-matter strong coupling fundamentally changes the nature of the molecular excitation and can lead to more efficient interactions with external probes.
In the second panel of \autoref{fig:intramolecular energy dipole transfer}, we report the photon coordinate computed for different quantum pictures: the QED-HF coherent state representation, the length gauge, and the velocity gauge.
That is, the computed quantities are
\begin{gather}
    \frac{\braket{b^\dagger + b}}{\sqrt{2\omega}} (t)\label{eq:coherent_state_photon_displacemet_field}\\
    \frac{\braket{b^\dagger + b - \sqrt{\frac{2}{\omega}}\;\bm{\lambda}\cdot \braket{\bm{d}}_{\text{QED-HF}}}}{\sqrt{2\omega}} (t)\label{eq:photon_displacemet_field}\\
    \frac{\braket{b^\dagger + b + \sqrt{\frac{2}{\omega}}\;\bm{\lambda}\cdot (\bm{d}-\braket{\bm{d}}_{\text{QED-HF}})}}{\sqrt{2\omega}} (t)\label{eq:photon_field}.
\end{gather}  
While for the $H_2$ system, the differences are less relevant, the permanent dipole moment of $\text C_4\text H_6\text O_3$ effectively shifts the photon coordinate values.
\autoref{fig:intramolecular energy dipole transfer} thus shows that the picture changes have a fundamental effect in determining the values of photonic observables, and care must be taken when reporting theoretical simulations.\cite{castagnola2024polaritonic}
It is also interesting to notice that the fast oscillations of the photonic observables have an opposite sign of the oscillations in the y dipole moment, that is, they are dephased by $\pi$. 

In addition, when the excited molecule transfers energy to the photon field, it can then be redistributed among all the states relevantly coupled to the optical device.
The cavity field thus favors an energy redistribution from the acidic moiety to the aldehyde group (long-range energy transfer) via the states I and III, and vice versa.
Notice, however, that these transfers of excitations are not equally feasible: state II has a larger transition dipole moment than I, and thus its coupling to the photon field is favored.
The cavity also promotes an excitation rearrangement of the aldehyde, transferring energy from state II to state I.
The intramolecular energy transfer between distant moieties of a large molecule is a chemically relevant potential application of electronic strong coupling.
We thus speculate that electronic and vibrational strong coupling could affect charge migration and excitation transfer, which should be further investigated.
In a sample with a large number of molecules, both inter- and intramolecular energy transfer would occur, and it might be challenging to separate the two effects.
However, single-molecule (or few-molecules) light-matter strong coupling has been achieved in plasmonic cavities, and the developed ideas apply equally to such devices.

\section{Conclusions}\label{sec: conclusion}

In this paper, we have developed a real-time coupled cluster theory with quantized electromagnetic fields (RT-QED-CC).
Our model can describe the real-time electron-photon dynamics of molecules coupled to optic devices such as microcavities or plasmonic structures, providing the time evolution of both matter and photon observables.
Moreover, the CC parametrization includes both electron-photon and short-range electron-electron correlation. 
The method is thus also suited to study the interplay between the intermolecular forces and (collective) polaritonic environment, which is suggested to induce local chemical effects.\cite{castagnola2024collective, biswas2024electronic, haugland2021intermolecular}
The \textit{ab initio} description of light \textit{and} matter can, in principle, be used to simulate the QED dynamics under both intense external fields (linear and nonlinear excitation regimes) and considerable light-matter coupling strengths (weak to ultrastrong coupling). 

We use the developed method to investigate the electron-photon dynamics of polariton-mediated energy transfer processes, which can lead to advances in technologies such as solar cells.
Our method allows for a microscopic description of the electron-photon dynamics, highlighting the light-matter interplay even on an attosecond scale.
The QED-CC wave function gives us access to the dipole moments and the photon coordinate, which provide a clear quantitative description of the system's evolution in the time domain.
The spatial visualization of the transport process is also obtained from the charge density displacement.
Our results highlight different time scales in the matter and photon observables and furthermore prove the relevance of the dark states in such polaritonic processes.
We discuss the requirements for photon-mediated energy transfer processes, which differ from the standard electronic F{\"o}rster or Dexter mechanisms, and argue that
such energy exchange processes can also be intramolecular, providing a channel alternative to molecular vibrations.
We thus speculate that experimental investigations of intramolecular charge and energy migration under electronic and vibrational strong coupling could provide novel applications of polaritonic chemistry.
Finally, we show that multiple classical pulses can alter the energy transfer dynamics and generate nontrivial dynamics of real photons inside the optical device.
While our method is focused on electronic strong coupling (ESC), we believe the developed concepts are valid also for vibrational strong coupling (VSC).\cite{xiang2020intermolecular, cao2022generalized, tibben2023molecular, li2021collective}

The developed method is well suited for supporting the study of experiments, including nonlinear optical processes,\cite{wang2021large, ge2021strongly, chervy2016high, malave2022real, xiang2019manipulating} quantum-light,\cite{dorfman2016nonlinear, schlawin2013two, ruggenthaler2018quantum, raymer2021entangled, schlawin2018entangled, rezus2012single, raimond2001manipulating} multi-photon, and transient absorption spectroscopies of molecules and molecular polaritons.\cite{gu2023cavity, delpo2020polariton, wang2023study, wang2018dynamics}
The study of nonlinear excitation regimes and pump-probe experiments in optical cavities will be investigated in future works, providing further insights into cavity-modified chemistry.

\section*{Supporting Information}

The Supporting Information includes the computational details of all the calculations, details on the RT-QED-CCSD-1 implementation in $e^\mathcal{T}$, and additional simulations of the systems under study.

\section*{Data avalability}
Videos of the energy transfer dynamics and output files are available in the following repository \doi{10.5281/zenodo.10813660}.

\section*{Acknowledgements}
M.C. acknowledges Andreas S. Skeidsvoll for insightful discussions.
M.C. and H.K. acknowledge funding from the European Research Council (ERC) under the European Union’s Horizon 2020 Research and Innovation Programme (grant agreement No. 101020016).
M. L. and H. K. acknowledge funding from the Department of Chemistry at the Norwegian University of Science and Technology (NTNU).
E.R. acknowledges funding from the European Research Council (ERC) under the European Union’s Horizon Europe Research and Innovation Programme (Grant n. ERC-StG-2021-101040197 - QED-SPIN).

\bibliography{lit.bib}

\providecommand{\latin}[1]{#1}
\makeatletter
\providecommand{\doi}
  {\begingroup\let\do\@makeother\dospecials
  \catcode`\{=1 \catcode`\}=2 \doi@aux}
\providecommand{\doi@aux}[1]{\endgroup\texttt{#1}}
\makeatother
\providecommand*\mcitethebibliography{\thebibliography}
\csname @ifundefined\endcsname{endmcitethebibliography}
  {\let\endmcitethebibliography\endthebibliography}{}
\begin{mcitethebibliography}{56}
\providecommand*\natexlab[1]{#1}
\providecommand*\mciteSetBstSublistMode[1]{}
\providecommand*\mciteSetBstMaxWidthForm[2]{}
\providecommand*\mciteBstWouldAddEndPuncttrue
  {\def\EndOfBibitem{\unskip.}}
\providecommand*\mciteBstWouldAddEndPunctfalse
  {\let\EndOfBibitem\relax}
\providecommand*\mciteSetBstMidEndSepPunct[3]{}
\providecommand*\mciteSetBstSublistLabelBeginEnd[3]{}
\providecommand*\EndOfBibitem{}
\mciteSetBstSublistMode{f}
\mciteSetBstMaxWidthForm{subitem}{(\alph{mcitesubitemcount})}
\mciteSetBstSublistLabelBeginEnd
  {\mcitemaxwidthsubitemform\space}
  {\relax}
  {\relax}

\bibitem[Hutchison \latin{et~al.}(2012)Hutchison, Schwartz, Genet, Devaux, and
  Ebbesen]{hutchison2012modifying}
Hutchison,~J.~A.; Schwartz,~T.; Genet,~C.; Devaux,~E.; Ebbesen,~T.~W. Modifying
  chemical landscapes by coupling to vacuum fields. \emph{Angewandte Chemie
  International Edition} \textbf{2012}, \emph{51}, 1592--1596\relax
\mciteBstWouldAddEndPuncttrue
\mciteSetBstMidEndSepPunct{\mcitedefaultmidpunct}
{\mcitedefaultendpunct}{\mcitedefaultseppunct}\relax
\EndOfBibitem
\bibitem[Munkhbat \latin{et~al.}(2018)Munkhbat, Wers{\"a}ll, Baranov,
  Antosiewicz, and Shegai]{munkhbat2018suppression}
Munkhbat,~B.; Wers{\"a}ll,~M.; Baranov,~D.~G.; Antosiewicz,~T.~J.; Shegai,~T.
  Suppression of photo-oxidation of organic chromophores by strong coupling to
  plasmonic nanoantennas. \emph{Science Advances} \textbf{2018}, \emph{4},
  eaas9552\relax
\mciteBstWouldAddEndPuncttrue
\mciteSetBstMidEndSepPunct{\mcitedefaultmidpunct}
{\mcitedefaultendpunct}{\mcitedefaultseppunct}\relax
\EndOfBibitem
\bibitem[Thomas \latin{et~al.}(2016)Thomas, George, Shalabney, Dryzhakov,
  Varma, Moran, Chervy, Zhong, Devaux, Genet, \latin{et~al.}
  others]{thomas2016ground}
Thomas,~A.; George,~J.; Shalabney,~A.; Dryzhakov,~M.; Varma,~S.~J.; Moran,~J.;
  Chervy,~T.; Zhong,~X.; Devaux,~E.; Genet,~C., \latin{et~al.}  Ground-state
  chemical reactivity under vibrational coupling to the vacuum electromagnetic
  field. \emph{Angewandte Chemie} \textbf{2016}, \emph{128}, 11634--11638\relax
\mciteBstWouldAddEndPuncttrue
\mciteSetBstMidEndSepPunct{\mcitedefaultmidpunct}
{\mcitedefaultendpunct}{\mcitedefaultseppunct}\relax
\EndOfBibitem
\bibitem[Lather \latin{et~al.}(2019)Lather, Bhatt, Thomas, Ebbesen, and
  George]{lather2019cavity}
Lather,~J.; Bhatt,~P.; Thomas,~A.; Ebbesen,~T.~W.; George,~J. Cavity catalysis
  by cooperative vibrational strong coupling of reactant and solvent molecules.
  \emph{Angewandte Chemie} \textbf{2019}, \emph{131}, 10745--10748\relax
\mciteBstWouldAddEndPuncttrue
\mciteSetBstMidEndSepPunct{\mcitedefaultmidpunct}
{\mcitedefaultendpunct}{\mcitedefaultseppunct}\relax
\EndOfBibitem
\bibitem[Thomas \latin{et~al.}(2019)Thomas, Lethuillier-Karl, Nagarajan,
  Vergauwe, George, Chervy, Shalabney, Devaux, Genet, Moran, \latin{et~al.}
  others]{thomas2019tilting}
Thomas,~A.; Lethuillier-Karl,~L.; Nagarajan,~K.; Vergauwe,~R.~M.; George,~J.;
  Chervy,~T.; Shalabney,~A.; Devaux,~E.; Genet,~C.; Moran,~J., \latin{et~al.}
  Tilting a ground-state reactivity landscape by vibrational strong coupling.
  \emph{Science} \textbf{2019}, \emph{363}, 615--619\relax
\mciteBstWouldAddEndPuncttrue
\mciteSetBstMidEndSepPunct{\mcitedefaultmidpunct}
{\mcitedefaultendpunct}{\mcitedefaultseppunct}\relax
\EndOfBibitem
\bibitem[Canaguier-Durand \latin{et~al.}(2013)Canaguier-Durand, Devaux, George,
  Pang, Hutchison, Schwartz, Genet, Wilhelms, Lehn, and
  Ebbesen]{canaguier2013thermodynamics}
Canaguier-Durand,~A.; Devaux,~E.; George,~J.; Pang,~Y.; Hutchison,~J.~A.;
  Schwartz,~T.; Genet,~C.; Wilhelms,~N.; Lehn,~J.-M.; Ebbesen,~T.~W.
  Thermodynamics of molecules strongly coupled to the vacuum field.
  \emph{Angewandte Chemie International Edition} \textbf{2013}, \emph{52},
  10533--10536\relax
\mciteBstWouldAddEndPuncttrue
\mciteSetBstMidEndSepPunct{\mcitedefaultmidpunct}
{\mcitedefaultendpunct}{\mcitedefaultseppunct}\relax
\EndOfBibitem
\bibitem[Sau \latin{et~al.}(2021)Sau, Nagarajan, Patrahau, Lethuillier-Karl,
  Vergauwe, Thomas, Moran, Genet, and Ebbesen]{sau2021modifying}
Sau,~A.; Nagarajan,~K.; Patrahau,~B.; Lethuillier-Karl,~L.; Vergauwe,~R.~M.;
  Thomas,~A.; Moran,~J.; Genet,~C.; Ebbesen,~T.~W. Modifying Woodward--Hoffmann
  stereoselectivity under vibrational strong coupling. \emph{Angewandte Chemie
  International Edition} \textbf{2021}, \emph{60}, 5712--5717\relax
\mciteBstWouldAddEndPuncttrue
\mciteSetBstMidEndSepPunct{\mcitedefaultmidpunct}
{\mcitedefaultendpunct}{\mcitedefaultseppunct}\relax
\EndOfBibitem
\bibitem[Hirai \latin{et~al.}(2020)Hirai, Takeda, Hutchison, and
  Uji-i]{hirai2020modulation}
Hirai,~K.; Takeda,~R.; Hutchison,~J.~A.; Uji-i,~H. Modulation of Prins
  cyclization by vibrational strong coupling. \emph{Angewandte Chemie}
  \textbf{2020}, \emph{132}, 5370--5373\relax
\mciteBstWouldAddEndPuncttrue
\mciteSetBstMidEndSepPunct{\mcitedefaultmidpunct}
{\mcitedefaultendpunct}{\mcitedefaultseppunct}\relax
\EndOfBibitem
\bibitem[Ahn \latin{et~al.}(2023)Ahn, Triana, Recabal, Herrera, and
  Simpkins]{ahn2023modification}
Ahn,~W.; Triana,~J.~F.; Recabal,~F.; Herrera,~F.; Simpkins,~B.~S. Modification
  of ground-state chemical reactivity via light--matter coherence in infrared
  cavities. \emph{Science} \textbf{2023}, \emph{380}, 1165--1168\relax
\mciteBstWouldAddEndPuncttrue
\mciteSetBstMidEndSepPunct{\mcitedefaultmidpunct}
{\mcitedefaultendpunct}{\mcitedefaultseppunct}\relax
\EndOfBibitem
\bibitem[Wang \latin{et~al.}(2018)Wang, Wang, Sun, Cerea, Toma, De~Angelis,
  Jin, Razzari, Cojoc, Catone, \latin{et~al.} others]{wang2018dynamics}
Wang,~H.; Wang,~H.-Y.; Sun,~H.-B.; Cerea,~A.; Toma,~A.; De~Angelis,~F.;
  Jin,~X.; Razzari,~L.; Cojoc,~D.; Catone,~D., \latin{et~al.}  Dynamics of
  strongly coupled hybrid states by transient absorption spectroscopy.
  \emph{Advanced Functional Materials} \textbf{2018}, \emph{28}, 1801761\relax
\mciteBstWouldAddEndPuncttrue
\mciteSetBstMidEndSepPunct{\mcitedefaultmidpunct}
{\mcitedefaultendpunct}{\mcitedefaultseppunct}\relax
\EndOfBibitem
\bibitem[Wang \latin{et~al.}(2023)Wang, Nagarajan, Kushida, Kulangara, Genet,
  and Ebbesen]{wang2023study}
Wang,~K.; Nagarajan,~K.; Kushida,~S.; Kulangara,~S.; Genet,~C.; Ebbesen,~T.
  Study of the selection rules of molecular polaritonic transitions by
  two-photon absorption spectroscopy. \textbf{2023}, \relax
\mciteBstWouldAddEndPunctfalse
\mciteSetBstMidEndSepPunct{\mcitedefaultmidpunct}
{}{\mcitedefaultseppunct}\relax
\EndOfBibitem
\bibitem[Georgiou \latin{et~al.}(2021)Georgiou, Jayaprakash, Othonos, and
  Lidzey]{georgiou2021ultralong}
Georgiou,~K.; Jayaprakash,~R.; Othonos,~A.; Lidzey,~D.~G. Ultralong-Range
  Polariton-Assisted Energy Transfer in Organic Microcavities. \emph{Angewandte
  Chemie} \textbf{2021}, \emph{133}, 16797--16803\relax
\mciteBstWouldAddEndPuncttrue
\mciteSetBstMidEndSepPunct{\mcitedefaultmidpunct}
{\mcitedefaultendpunct}{\mcitedefaultseppunct}\relax
\EndOfBibitem
\bibitem[Zhong \latin{et~al.}(2017)Zhong, Chervy, Zhang, Thomas, George, Genet,
  Hutchison, and Ebbesen]{zhong2017energy}
Zhong,~X.; Chervy,~T.; Zhang,~L.; Thomas,~A.; George,~J.; Genet,~C.;
  Hutchison,~J.~A.; Ebbesen,~T.~W. Energy transfer between spatially separated
  entangled molecules. \emph{Angewandte Chemie} \textbf{2017}, \emph{129},
  9162--9166\relax
\mciteBstWouldAddEndPuncttrue
\mciteSetBstMidEndSepPunct{\mcitedefaultmidpunct}
{\mcitedefaultendpunct}{\mcitedefaultseppunct}\relax
\EndOfBibitem
\bibitem[Zhong \latin{et~al.}(2016)Zhong, Chervy, Wang, George, Thomas,
  Hutchison, Devaux, Genet, and Ebbesen]{zhong2016non}
Zhong,~X.; Chervy,~T.; Wang,~S.; George,~J.; Thomas,~A.; Hutchison,~J.~A.;
  Devaux,~E.; Genet,~C.; Ebbesen,~T.~W. Non-Radiative Energy Transfer Mediated
  by Hybrid Light-Matter States. \emph{Angewandte Chemie} \textbf{2016},
  \emph{128}, 6310--6314\relax
\mciteBstWouldAddEndPuncttrue
\mciteSetBstMidEndSepPunct{\mcitedefaultmidpunct}
{\mcitedefaultendpunct}{\mcitedefaultseppunct}\relax
\EndOfBibitem
\bibitem[Cargioli \latin{et~al.}(2024)Cargioli, Lednev, Lavista, Camposeo,
  Sassella, Pisignano, Tredicucci, Garcia-Vidal, Feist, and
  Persano]{cargioli2024active}
Cargioli,~A.; Lednev,~M.; Lavista,~L.; Camposeo,~A.; Sassella,~A.;
  Pisignano,~D.; Tredicucci,~A.; Garcia-Vidal,~F.~J.; Feist,~J.; Persano,~L.
  Active control of polariton-enabled long-range energy transfer.
  \emph{Nanophotonics} \textbf{2024}, \relax
\mciteBstWouldAddEndPunctfalse
\mciteSetBstMidEndSepPunct{\mcitedefaultmidpunct}
{}{\mcitedefaultseppunct}\relax
\EndOfBibitem
\bibitem[Coles \latin{et~al.}(2014)Coles, Somaschi, Michetti, Clark,
  Lagoudakis, Savvidis, and Lidzey]{coles2014polariton}
Coles,~D.~M.; Somaschi,~N.; Michetti,~P.; Clark,~C.; Lagoudakis,~P.~G.;
  Savvidis,~P.~G.; Lidzey,~D.~G. Polariton-mediated energy transfer between
  organic dyes in a strongly coupled optical microcavity. \emph{Nature
  materials} \textbf{2014}, \emph{13}, 712--719\relax
\mciteBstWouldAddEndPuncttrue
\mciteSetBstMidEndSepPunct{\mcitedefaultmidpunct}
{\mcitedefaultendpunct}{\mcitedefaultseppunct}\relax
\EndOfBibitem
\bibitem[Sch{\"a}fer \latin{et~al.}(2019)Sch{\"a}fer, Ruggenthaler, Appel, and
  Rubio]{schafer2019modification}
Sch{\"a}fer,~C.; Ruggenthaler,~M.; Appel,~H.; Rubio,~A. Modification of
  excitation and charge transfer in cavity quantum-electrodynamical chemistry.
  \emph{Proceedings of the National Academy of Sciences} \textbf{2019},
  \emph{116}, 4883--4892\relax
\mciteBstWouldAddEndPuncttrue
\mciteSetBstMidEndSepPunct{\mcitedefaultmidpunct}
{\mcitedefaultendpunct}{\mcitedefaultseppunct}\relax
\EndOfBibitem
\bibitem[Xiang \latin{et~al.}(2020)Xiang, Ribeiro, Du, Chen, Yang, Wang,
  Yuen-Zhou, and Xiong]{xiang2020intermolecular}
Xiang,~B.; Ribeiro,~R.~F.; Du,~M.; Chen,~L.; Yang,~Z.; Wang,~J.; Yuen-Zhou,~J.;
  Xiong,~W. Intermolecular vibrational energy transfer enabled by microcavity
  strong light--matter coupling. \emph{Science} \textbf{2020}, \emph{368},
  665--667\relax
\mciteBstWouldAddEndPuncttrue
\mciteSetBstMidEndSepPunct{\mcitedefaultmidpunct}
{\mcitedefaultendpunct}{\mcitedefaultseppunct}\relax
\EndOfBibitem
\bibitem[Wang \latin{et~al.}(2021)Wang, Seidel, Nagarajan, Chervy, Genet, and
  Ebbesen]{wang2021large}
Wang,~K.; Seidel,~M.; Nagarajan,~K.; Chervy,~T.; Genet,~C.; Ebbesen,~T. Large
  optical nonlinearity enhancement under electronic strong coupling.
  \emph{Nature Communications} \textbf{2021}, \emph{12}, 1486\relax
\mciteBstWouldAddEndPuncttrue
\mciteSetBstMidEndSepPunct{\mcitedefaultmidpunct}
{\mcitedefaultendpunct}{\mcitedefaultseppunct}\relax
\EndOfBibitem
\bibitem[Ge \latin{et~al.}(2021)Ge, Han, and Xu]{ge2021strongly}
Ge,~F.; Han,~X.; Xu,~J. Strongly coupled systems for nonlinear optics.
  \emph{Laser \& Photonics Reviews} \textbf{2021}, \emph{15}, 2000514\relax
\mciteBstWouldAddEndPuncttrue
\mciteSetBstMidEndSepPunct{\mcitedefaultmidpunct}
{\mcitedefaultendpunct}{\mcitedefaultseppunct}\relax
\EndOfBibitem
\bibitem[Chervy \latin{et~al.}(2016)Chervy, Xu, Duan, Wang, Mager, Frerejean,
  Munninghoff, Tinnemans, Hutchison, Genet, \latin{et~al.}
  others]{chervy2016high}
Chervy,~T.; Xu,~J.; Duan,~Y.; Wang,~C.; Mager,~L.; Frerejean,~M.;
  Munninghoff,~J.~A.; Tinnemans,~P.; Hutchison,~J.~A.; Genet,~C.,
  \latin{et~al.}  High-efficiency second-harmonic generation from hybrid
  light-matter states. \emph{Nano letters} \textbf{2016}, \emph{16},
  7352--7356\relax
\mciteBstWouldAddEndPuncttrue
\mciteSetBstMidEndSepPunct{\mcitedefaultmidpunct}
{\mcitedefaultendpunct}{\mcitedefaultseppunct}\relax
\EndOfBibitem
\bibitem[Malave \latin{et~al.}(2022)Malave, Ahrens, Pitagora, Covington, and
  Varga]{malave2022real}
Malave,~J.; Ahrens,~A.; Pitagora,~D.; Covington,~C.; Varga,~K. Real-space,
  real-time approach to quantum-electrodynamical time-dependent density
  functional theory. \emph{The Journal of chemical physics} \textbf{2022},
  \emph{157}\relax
\mciteBstWouldAddEndPuncttrue
\mciteSetBstMidEndSepPunct{\mcitedefaultmidpunct}
{\mcitedefaultendpunct}{\mcitedefaultseppunct}\relax
\EndOfBibitem
\bibitem[Xiang \latin{et~al.}(2019)Xiang, Ribeiro, Li, Dunkelberger, Simpkins,
  Yuen-Zhou, and Xiong]{xiang2019manipulating}
Xiang,~B.; Ribeiro,~R.~F.; Li,~Y.; Dunkelberger,~A.~D.; Simpkins,~B.~B.;
  Yuen-Zhou,~J.; Xiong,~W. Manipulating optical nonlinearities of molecular
  polaritons by delocalization. \emph{Science advances} \textbf{2019},
  \emph{5}, eaax5196\relax
\mciteBstWouldAddEndPuncttrue
\mciteSetBstMidEndSepPunct{\mcitedefaultmidpunct}
{\mcitedefaultendpunct}{\mcitedefaultseppunct}\relax
\EndOfBibitem
\bibitem[Haugland \latin{et~al.}(2020)Haugland, Ronca, Kj{\o}nstad, Rubio, and
  Koch]{haugland2020coupled}
Haugland,~T.~S.; Ronca,~E.; Kj{\o}nstad,~E.~F.; Rubio,~A.; Koch,~H. Coupled
  cluster theory for molecular polaritons: Changing ground and excited states.
  \emph{Physical Review X} \textbf{2020}, \emph{10}, 041043\relax
\mciteBstWouldAddEndPuncttrue
\mciteSetBstMidEndSepPunct{\mcitedefaultmidpunct}
{\mcitedefaultendpunct}{\mcitedefaultseppunct}\relax
\EndOfBibitem
\bibitem[Helgaker \latin{et~al.}(2013)Helgaker, Jorgensen, and
  Olsen]{helgaker2013molecular}
Helgaker,~T.; Jorgensen,~P.; Olsen,~J. \emph{Molecular electronic-structure
  theory}; John Wiley \& Sons, 2013\relax
\mciteBstWouldAddEndPuncttrue
\mciteSetBstMidEndSepPunct{\mcitedefaultmidpunct}
{\mcitedefaultendpunct}{\mcitedefaultseppunct}\relax
\EndOfBibitem
\bibitem[Dorfman \latin{et~al.}(2016)Dorfman, Schlawin, and
  Mukamel]{dorfman2016nonlinear}
Dorfman,~K.~E.; Schlawin,~F.; Mukamel,~S. Nonlinear optical signals and
  spectroscopy with quantum light. \emph{Reviews of Modern Physics}
  \textbf{2016}, \emph{88}, 045008\relax
\mciteBstWouldAddEndPuncttrue
\mciteSetBstMidEndSepPunct{\mcitedefaultmidpunct}
{\mcitedefaultendpunct}{\mcitedefaultseppunct}\relax
\EndOfBibitem
\bibitem[Schlawin and Mukamel(2013)Schlawin, and Mukamel]{schlawin2013two}
Schlawin,~F.; Mukamel,~S. Two-photon spectroscopy of excitons with entangled
  photons. \emph{The Journal of chemical physics} \textbf{2013},
  \emph{139}\relax
\mciteBstWouldAddEndPuncttrue
\mciteSetBstMidEndSepPunct{\mcitedefaultmidpunct}
{\mcitedefaultendpunct}{\mcitedefaultseppunct}\relax
\EndOfBibitem
\bibitem[Ruggenthaler \latin{et~al.}(2018)Ruggenthaler, Tancogne-Dejean, Flick,
  Appel, and Rubio]{ruggenthaler2018quantum}
Ruggenthaler,~M.; Tancogne-Dejean,~N.; Flick,~J.; Appel,~H.; Rubio,~A. From a
  quantum-electrodynamical light--matter description to novel spectroscopies.
  \emph{Nature Reviews Chemistry} \textbf{2018}, \emph{2}, 1--16\relax
\mciteBstWouldAddEndPuncttrue
\mciteSetBstMidEndSepPunct{\mcitedefaultmidpunct}
{\mcitedefaultendpunct}{\mcitedefaultseppunct}\relax
\EndOfBibitem
\bibitem[Raymer \latin{et~al.}(2021)Raymer, Landes, and
  Marcus]{raymer2021entangled}
Raymer,~M.~G.; Landes,~T.; Marcus,~A.~H. Entangled two-photon absorption by
  atoms and molecules: A quantum optics tutorial. \emph{The Journal of Chemical
  Physics} \textbf{2021}, \emph{155}\relax
\mciteBstWouldAddEndPuncttrue
\mciteSetBstMidEndSepPunct{\mcitedefaultmidpunct}
{\mcitedefaultendpunct}{\mcitedefaultseppunct}\relax
\EndOfBibitem
\bibitem[Schlawin \latin{et~al.}(2018)Schlawin, Dorfman, and
  Mukamel]{schlawin2018entangled}
Schlawin,~F.; Dorfman,~K.~E.; Mukamel,~S. Entangled two-photon absorption
  spectroscopy. \emph{Accounts of chemical research} \textbf{2018}, \emph{51},
  2207--2214\relax
\mciteBstWouldAddEndPuncttrue
\mciteSetBstMidEndSepPunct{\mcitedefaultmidpunct}
{\mcitedefaultendpunct}{\mcitedefaultseppunct}\relax
\EndOfBibitem
\bibitem[Rezus \latin{et~al.}(2012)Rezus, Walt, Lettow, Renn, Zumofen,
  G{\"o}tzinger, and Sandoghdar]{rezus2012single}
Rezus,~Y.; Walt,~S.; Lettow,~R.; Renn,~A.; Zumofen,~G.; G{\"o}tzinger,~S.;
  Sandoghdar,~V. Single-photon spectroscopy of a single molecule.
  \emph{Physical review letters} \textbf{2012}, \emph{108}, 093601\relax
\mciteBstWouldAddEndPuncttrue
\mciteSetBstMidEndSepPunct{\mcitedefaultmidpunct}
{\mcitedefaultendpunct}{\mcitedefaultseppunct}\relax
\EndOfBibitem
\bibitem[Raimond \latin{et~al.}(2001)Raimond, Brune, and
  Haroche]{raimond2001manipulating}
Raimond,~J.-M.; Brune,~M.; Haroche,~S. Manipulating quantum entanglement with
  atoms and photons in a cavity. \emph{Reviews of Modern Physics}
  \textbf{2001}, \emph{73}, 565\relax
\mciteBstWouldAddEndPuncttrue
\mciteSetBstMidEndSepPunct{\mcitedefaultmidpunct}
{\mcitedefaultendpunct}{\mcitedefaultseppunct}\relax
\EndOfBibitem
\bibitem[Gu \latin{et~al.}(2023)Gu, Gu, Chernyak, and Mukamel]{gu2023cavity}
Gu,~B.; Gu,~Y.; Chernyak,~V.~Y.; Mukamel,~S. Cavity Control of Molecular
  Spectroscopy and Photophysics. \emph{Accounts of Chemical Research}
  \textbf{2023}, \emph{56}, 2753--2762\relax
\mciteBstWouldAddEndPuncttrue
\mciteSetBstMidEndSepPunct{\mcitedefaultmidpunct}
{\mcitedefaultendpunct}{\mcitedefaultseppunct}\relax
\EndOfBibitem
\bibitem[DelPo \latin{et~al.}(2020)DelPo, Kudisch, Park, Khan, Fassioli,
  Fausti, Rand, and Scholes]{delpo2020polariton}
DelPo,~C.~A.; Kudisch,~B.; Park,~K.~H.; Khan,~S.-U.-Z.; Fassioli,~F.;
  Fausti,~D.; Rand,~B.~P.; Scholes,~G.~D. Polariton transitions in femtosecond
  transient absorption studies of ultrastrong light--molecule coupling.
  \emph{The journal of physical chemistry letters} \textbf{2020}, \emph{11},
  2667--2674\relax
\mciteBstWouldAddEndPuncttrue
\mciteSetBstMidEndSepPunct{\mcitedefaultmidpunct}
{\mcitedefaultendpunct}{\mcitedefaultseppunct}\relax
\EndOfBibitem
\bibitem[Cao(2022)]{cao2022generalized}
Cao,~J. Generalized resonance energy transfer theory: Applications to
  vibrational energy flow in optical cavities. \emph{The Journal of Physical
  Chemistry Letters} \textbf{2022}, \emph{13}, 10943--10951\relax
\mciteBstWouldAddEndPuncttrue
\mciteSetBstMidEndSepPunct{\mcitedefaultmidpunct}
{\mcitedefaultendpunct}{\mcitedefaultseppunct}\relax
\EndOfBibitem
\bibitem[Tibben \latin{et~al.}(2023)Tibben, Bonin, Cho, Lakhwani, Hutchison,
  and G{\'o}mez]{tibben2023molecular}
Tibben,~D.~J.; Bonin,~G.~O.; Cho,~I.; Lakhwani,~G.; Hutchison,~J.;
  G{\'o}mez,~D.~E. Molecular energy transfer under the strong light--matter
  interaction regime. \emph{Chemical Reviews} \textbf{2023}, \emph{123},
  8044--8068\relax
\mciteBstWouldAddEndPuncttrue
\mciteSetBstMidEndSepPunct{\mcitedefaultmidpunct}
{\mcitedefaultendpunct}{\mcitedefaultseppunct}\relax
\EndOfBibitem
\bibitem[Li \latin{et~al.}(2021)Li, Nitzan, and Subotnik]{li2021collective}
Li,~T.~E.; Nitzan,~A.; Subotnik,~J.~E. Collective vibrational strong coupling
  effects on molecular vibrational relaxation and energy transfer: Numerical
  insights via cavity molecular dynamics simulations. \emph{Angewandte Chemie}
  \textbf{2021}, \emph{133}, 15661--15668\relax
\mciteBstWouldAddEndPuncttrue
\mciteSetBstMidEndSepPunct{\mcitedefaultmidpunct}
{\mcitedefaultendpunct}{\mcitedefaultseppunct}\relax
\EndOfBibitem
\bibitem[Ruggenthaler \latin{et~al.}(2023)Ruggenthaler, Sidler, and
  Rubio]{ruggenthaler2023understanding}
Ruggenthaler,~M.; Sidler,~D.; Rubio,~A. Understanding polaritonic chemistry
  from ab initio quantum electrodynamics. \emph{Chemical Reviews}
  \textbf{2023}, \emph{123}, 11191--11229\relax
\mciteBstWouldAddEndPuncttrue
\mciteSetBstMidEndSepPunct{\mcitedefaultmidpunct}
{\mcitedefaultendpunct}{\mcitedefaultseppunct}\relax
\EndOfBibitem
\bibitem[Castagnola \latin{et~al.}(2024)Castagnola, Riso, Barlini, Ronca, and
  Koch]{castagnola2024polaritonic}
Castagnola,~M.; Riso,~R.~R.; Barlini,~A.; Ronca,~E.; Koch,~H. Polaritonic
  response theory for exact and approximate wave functions. \emph{Wiley
  Interdisciplinary Reviews: Computational Molecular Science} \textbf{2024},
  \emph{14}, e1684\relax
\mciteBstWouldAddEndPuncttrue
\mciteSetBstMidEndSepPunct{\mcitedefaultmidpunct}
{\mcitedefaultendpunct}{\mcitedefaultseppunct}\relax
\EndOfBibitem
\bibitem[Pedersen and Koch(1997)Pedersen, and Koch]{pedersen1997coupled}
Pedersen,~T.~B.; Koch,~H. Coupled cluster response functions revisited.
  \emph{The Journal of chemical physics} \textbf{1997}, \emph{106},
  8059--8072\relax
\mciteBstWouldAddEndPuncttrue
\mciteSetBstMidEndSepPunct{\mcitedefaultmidpunct}
{\mcitedefaultendpunct}{\mcitedefaultseppunct}\relax
\EndOfBibitem
\bibitem[Koch and Jorgensen(1990)Koch, and Jorgensen]{koch1990coupled}
Koch,~H.; Jorgensen,~P. Coupled cluster response functions. \emph{The Journal
  of chemical physics} \textbf{1990}, \emph{93}, 3333--3344\relax
\mciteBstWouldAddEndPuncttrue
\mciteSetBstMidEndSepPunct{\mcitedefaultmidpunct}
{\mcitedefaultendpunct}{\mcitedefaultseppunct}\relax
\EndOfBibitem
\bibitem[Sverdrup~Ofstad \latin{et~al.}(2023)Sverdrup~Ofstad, Aurbakken,
  Sigmundson~Sch{\o}yen, Kristiansen, Kvaal, and Pedersen]{sverdrup2023time}
Sverdrup~Ofstad,~B.; Aurbakken,~E.; Sigmundson~Sch{\o}yen,~{\O}.;
  Kristiansen,~H.~E.; Kvaal,~S.; Pedersen,~T.~B. Time-dependent coupled-cluster
  theory. \emph{Wiley Interdisciplinary Reviews: Computational Molecular
  Science} \textbf{2023}, \emph{13}, e1666\relax
\mciteBstWouldAddEndPuncttrue
\mciteSetBstMidEndSepPunct{\mcitedefaultmidpunct}
{\mcitedefaultendpunct}{\mcitedefaultseppunct}\relax
\EndOfBibitem
\bibitem[Skeidsvoll \latin{et~al.}(2020)Skeidsvoll, Balbi, and
  Koch]{skeidsvoll2020time}
Skeidsvoll,~A.~S.; Balbi,~A.; Koch,~H. Time-dependent coupled-cluster theory
  for ultrafast transient-absorption spectroscopy. \emph{Physical Review A}
  \textbf{2020}, \emph{102}, 023115\relax
\mciteBstWouldAddEndPuncttrue
\mciteSetBstMidEndSepPunct{\mcitedefaultmidpunct}
{\mcitedefaultendpunct}{\mcitedefaultseppunct}\relax
\EndOfBibitem
\bibitem[Skeidsvoll \latin{et~al.}(2022)Skeidsvoll, Moitra, Balbi, Paul,
  Coriani, and Koch]{skeidsvoll2022simulating}
Skeidsvoll,~A.~S.; Moitra,~T.; Balbi,~A.; Paul,~A.~C.; Coriani,~S.; Koch,~H.
  Simulating weak-field attosecond processes with a Lanczos reduced basis
  approach to time-dependent equation-of-motion coupled-cluster theory.
  \emph{Physical Review A} \textbf{2022}, \emph{105}, 023103\relax
\mciteBstWouldAddEndPuncttrue
\mciteSetBstMidEndSepPunct{\mcitedefaultmidpunct}
{\mcitedefaultendpunct}{\mcitedefaultseppunct}\relax
\EndOfBibitem
\bibitem[Nascimento and DePrince(2019)Nascimento, and
  DePrince]{nascimento2019general}
Nascimento,~D.~R.; DePrince,~A.~E. A general time-domain formulation of
  equation-of-motion coupled-cluster theory for linear spectroscopy. \emph{The
  Journal of Chemical Physics} \textbf{2019}, \emph{151}\relax
\mciteBstWouldAddEndPuncttrue
\mciteSetBstMidEndSepPunct{\mcitedefaultmidpunct}
{\mcitedefaultendpunct}{\mcitedefaultseppunct}\relax
\EndOfBibitem
\bibitem[Pedersen and Kvaal(2019)Pedersen, and Kvaal]{pedersen2019symplectic}
Pedersen,~T.~B.; Kvaal,~S. Symplectic integration and physical interpretation
  of time-dependent coupled-cluster theory. \emph{The Journal of chemical
  physics} \textbf{2019}, \emph{150}\relax
\mciteBstWouldAddEndPuncttrue
\mciteSetBstMidEndSepPunct{\mcitedefaultmidpunct}
{\mcitedefaultendpunct}{\mcitedefaultseppunct}\relax
\EndOfBibitem
\bibitem[Huber and Klamroth(2011)Huber, and Klamroth]{huber2011explicitly}
Huber,~C.; Klamroth,~T. Explicitly time-dependent coupled cluster singles
  doubles calculations of laser-driven many-electron dynamics. \emph{The
  Journal of chemical physics} \textbf{2011}, \emph{134}\relax
\mciteBstWouldAddEndPuncttrue
\mciteSetBstMidEndSepPunct{\mcitedefaultmidpunct}
{\mcitedefaultendpunct}{\mcitedefaultseppunct}\relax
\EndOfBibitem
\bibitem[Aurbakken \latin{et~al.}(2023)Aurbakken, Kristiansen, Kvaal, Pedersen,
  \latin{et~al.} others]{aurbakken2023time}
Aurbakken,~E.; Kristiansen,~H.; Kvaal,~S.; Pedersen,~T., \latin{et~al.}
  Time-dependent coupled-cluster theory. \emph{Wiley Interdiscip. Rev.: Comput.
  Mol. Sci.} \textbf{2023}, \emph{13}, e1666\relax
\mciteBstWouldAddEndPuncttrue
\mciteSetBstMidEndSepPunct{\mcitedefaultmidpunct}
{\mcitedefaultendpunct}{\mcitedefaultseppunct}\relax
\EndOfBibitem
\bibitem[Schlawin(2017)]{schlawin2017entangled}
Schlawin,~F. Entangled photon spectroscopy. \emph{Journal of Physics B: Atomic,
  Molecular and Optical Physics} \textbf{2017}, \emph{50}, 203001\relax
\mciteBstWouldAddEndPuncttrue
\mciteSetBstMidEndSepPunct{\mcitedefaultmidpunct}
{\mcitedefaultendpunct}{\mcitedefaultseppunct}\relax
\EndOfBibitem
\bibitem[Folkestad \latin{et~al.}(2020)Folkestad, Kj{\o}nstad, Myhre, Andersen,
  Balbi, Coriani, Giovannini, Goletto, Haugland, Hutcheson, \latin{et~al.}
  others]{folkestad2020t}
Folkestad,~S.~D.; Kj{\o}nstad,~E.~F.; Myhre,~R.~H.; Andersen,~J.~H.; Balbi,~A.;
  Coriani,~S.; Giovannini,~T.; Goletto,~L.; Haugland,~T.~S.; Hutcheson,~A.,
  \latin{et~al.}  e T 1.0: An open source electronic structure program with
  emphasis on coupled cluster and multilevel methods. \emph{The Journal of
  Chemical Physics} \textbf{2020}, \emph{152}, 184103\relax
\mciteBstWouldAddEndPuncttrue
\mciteSetBstMidEndSepPunct{\mcitedefaultmidpunct}
{\mcitedefaultendpunct}{\mcitedefaultseppunct}\relax
\EndOfBibitem
\bibitem[Scholes(2003)]{scholes2003long}
Scholes,~G.~D. Long-range resonance energy transfer in molecular systems.
  \emph{Annual review of physical chemistry} \textbf{2003}, \emph{54},
  57--87\relax
\mciteBstWouldAddEndPuncttrue
\mciteSetBstMidEndSepPunct{\mcitedefaultmidpunct}
{\mcitedefaultendpunct}{\mcitedefaultseppunct}\relax
\EndOfBibitem
\bibitem[Skeidsvoll and Koch(2023)Skeidsvoll, and
  Koch]{skeidsvoll2023comparing}
Skeidsvoll,~A.~S.; Koch,~H. Comparing real-time coupled-cluster methods through
  simulation of collective Rabi oscillations. \emph{Physical Review A}
  \textbf{2023}, \emph{108}, 033116\relax
\mciteBstWouldAddEndPuncttrue
\mciteSetBstMidEndSepPunct{\mcitedefaultmidpunct}
{\mcitedefaultendpunct}{\mcitedefaultseppunct}\relax
\EndOfBibitem
\bibitem[Castagnola \latin{et~al.}(2024)Castagnola, Haugland, Ronca, Koch, and
  Sch{\"a}fer]{castagnola2024collective}
Castagnola,~M.; Haugland,~T.~S.; Ronca,~E.; Koch,~H.; Sch{\"a}fer,~C.
  Collective strong coupling modifies aggregation and solvation. \emph{The
  Journal of Physical Chemistry Letters} \textbf{2024}, \emph{15},
  1428--1434\relax
\mciteBstWouldAddEndPuncttrue
\mciteSetBstMidEndSepPunct{\mcitedefaultmidpunct}
{\mcitedefaultendpunct}{\mcitedefaultseppunct}\relax
\EndOfBibitem
\bibitem[Biswas \latin{et~al.}(2024)Biswas, Mondal, Chandrasekharan, Singh, and
  Thomas]{biswas2024electronic}
Biswas,~S.; Mondal,~M.; Chandrasekharan,~G.; Singh,~A.; Thomas,~A. Electronic
  Strong Coupling Modifies the Ground-state Intermolecular Interactions in
  Chlorin Thin Films. \textbf{2024}, \relax
\mciteBstWouldAddEndPunctfalse
\mciteSetBstMidEndSepPunct{\mcitedefaultmidpunct}
{}{\mcitedefaultseppunct}\relax
\EndOfBibitem
\bibitem[Haugland \latin{et~al.}(2021)Haugland, Sch{\"a}fer, Ronca, Rubio, and
  Koch]{haugland2021intermolecular}
Haugland,~T.~S.; Sch{\"a}fer,~C.; Ronca,~E.; Rubio,~A.; Koch,~H. Intermolecular
  interactions in optical cavities: An ab initio QED study. \emph{The Journal
  of Chemical Physics} \textbf{2021}, \emph{154}, 094113\relax
\mciteBstWouldAddEndPuncttrue
\mciteSetBstMidEndSepPunct{\mcitedefaultmidpunct}
{\mcitedefaultendpunct}{\mcitedefaultseppunct}\relax
\EndOfBibitem
\end{mcitethebibliography}

\cleardoublepage
\includepdf[pages=-]{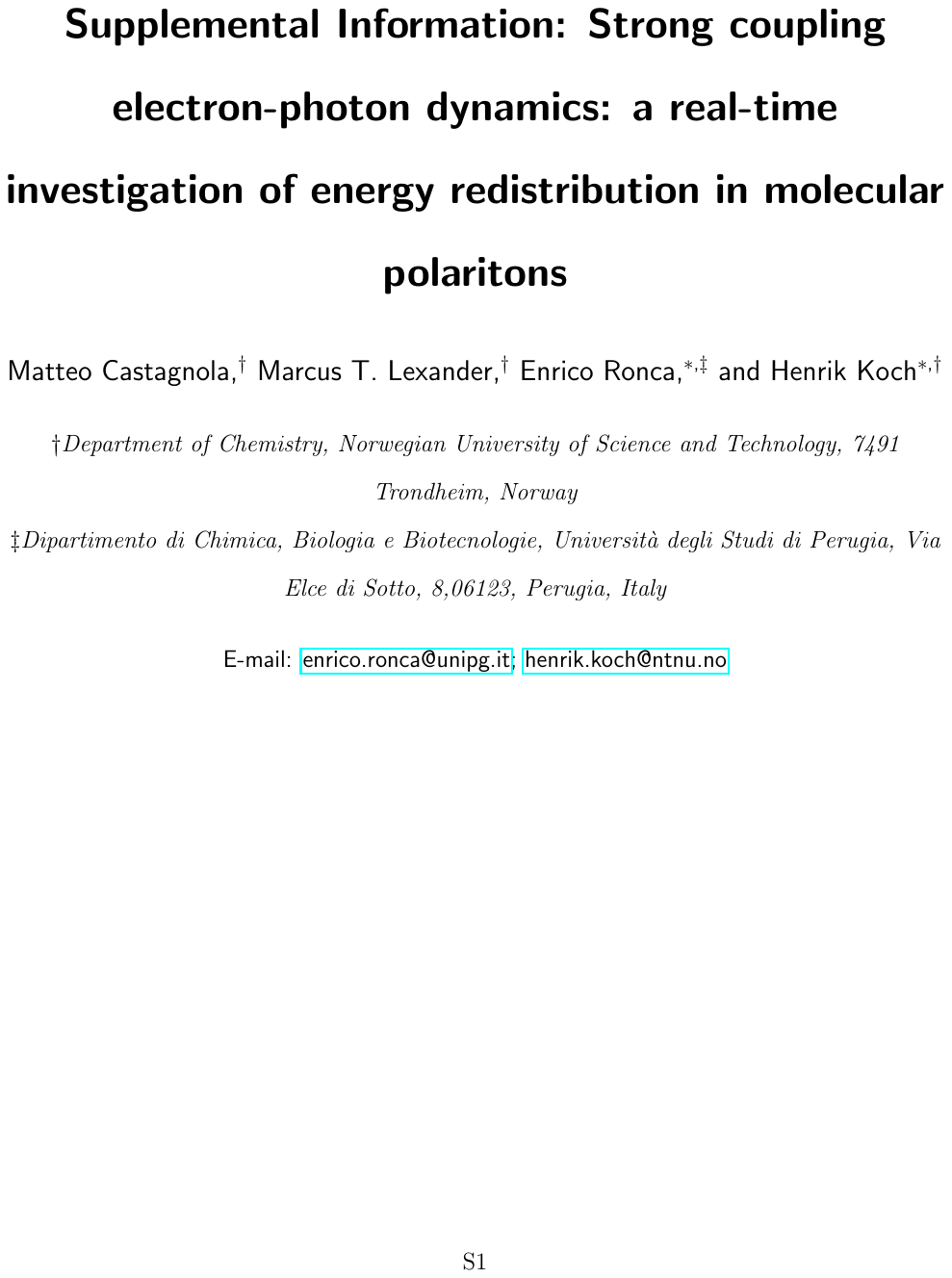}

\end{document}